\documentclass[twocolumn]{article}

\usepackage{arxiv}
\usepackage[utf8]{inputenc}             
\usepackage[T1]{fontenc}                
\usepackage{hyperref}                   
\usepackage{url}                        
\usepackage{svg}                        
\usepackage{booktabs}                   
\usepackage{amsfonts}                   
\usepackage{amsmath}                    
\usepackage{braket}                     
\usepackage{nicefrac}                   
\usepackage[letterspace=24]{microtype}  
\usepackage[multiple]{footmisc}         
\usepackage[labelfont=bf]{caption}      
\usepackage{graphicx}                   
\usepackage{cuted}                      
\usepackage[framemethod=tikz]{mdframed} 
\usetikzlibrary{shadows}
\usepackage{tabularx}                   
\usepackage{rotating}                   
\usepackage{chngpage}                   
\usepackage{enumitem}                   
\usepackage[font=small]{caption}        

\graphicspath{{./figures/}}             
\definecolor{box-color-inner}{cmyk}{0.38, 0.16, 0.0, 0.0}
\definecolor{box-color-shadow}{cmyk}{0.0, 0.2, 0.0, 0.04}

\mdfsetup{
    shadow              = true,
    shadowsize          = 6pt,
    shadowcolor         = box-color-shadow,
    backgroundcolor     = box-color-inner,
    linewidth           = 0pt,
    rightmargin         = 6pt,
    innertopmargin      = 12pt,
    innerbottommargin   = 12pt,
    innerleftmargin     = 12pt,
    innerrightmargin    = 12pt
}

\title{Biology and Medicine in the Landscape of Quantum Advantages}

\author{ 
    \href{https://orcid.org/0000-0001-5122-7879}{\includegraphics[scale=0.06]{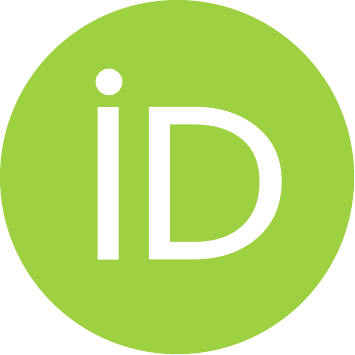}\hspace{1mm}Benjamin A. ~Cordier} \inst{1},
	\href{https://orcid.org/0000-0001-8510-8480}{\includegraphics[scale=0.06]{icons/orcid.pdf}\hspace{1mm}Nicolas P. D. Sawaya} \inst{2}, 
	\href{https://orcid.org/0000-0002-5579-451X}{\includegraphics[scale=0.06]{icons/orcid.pdf}\hspace{1mm}Gian Giacomo Guerreschi} \inst{2}, 
	\href{https://orcid.org/0000-0001-8333-6607}{\includegraphics[scale=0.06]{icons/orcid.pdf}\hspace{1mm}Shannon K. McWeeney} \inst{1,3,4}
}

\institute{
    Department of Medical Informatics \& Clinical Epidemiology, Oregon Health \& Science University, Portland, OR 97202, USA
    \and
    Intel Labs, Santa Clara, CA 95054, USA
    \and
    Knight Cancer Institute, Oregon Health \& Science University, Portland, OR 97202, USA
    \and Oregon Clinical and Translational Research Institute, Oregon Health \& Science University, Portland, OR 97202, USA
}

\contact{cordier@ohsu.edu}

\begin{document}

\tikzset{every shadow/.style={opacity=1}}

\twocolumn[
  \begin{@twocolumnfalse}
    \maketitle
    \begin{abstract}
    Quantum computing holds significant potential for applications in biology and medicine, spanning from the simulation of biomolecules to machine learning approaches for subtyping cancers on the basis of clinical features. This potential is encapsulated by the concept of a quantum advantage, which is typically contingent on a reduction in the consumption of a computational resource, such as time, space, or data. Here, we distill the concept of a quantum advantage into a simple framework that we hope will aid researchers in biology and medicine pursuing the development of quantum applications. We then apply this framework to a wide variety of computational problems relevant to these domains in an effort to i) assess the potential of quantum advantages in specific application areas and ii) identify gaps that may be addressed with novel quantum approaches. Bearing in mind the rapid pace of change in the fields of quantum computing and classical algorithms, we aim to provide an extensive survey of applications in biology and medicine that may lead to practical quantum advantages.
    \end{abstract}
    \vspace{8pt}
  \end{@twocolumnfalse}
]

\section{Introduction}
\label{sec:Section-I}

\subsection{Quantum computing: past to present}
\label{sec:Section-aboqc}

The notion that a quantum computer may be more powerful than a classical computer was first conceived some forty years ago in the context of simulating physical systems \cite{Mani80, Beni80, Feyn82}. Theoretical models of quantum computers quickly followed \cite{Deut85, Deut89, Yao93}. Then, in 1993, 11 years after Feynman’s talk on simulating physics \cite{Feyn82}, early formal evidence that a universal quantum computer may be more powerful than its classical counterpart arrived with proof of a superpolynomial quantum advantage on an artificial problem, recursive Fourier Sampling \cite{BV93}. 

This was shortly followed by the development of the first quantum algorithm for a practical problem, prime factorization, by Peter Shor in 1994 – also yielding a superpolynomial advantage \cite{Shor94}. \textrm{Shor’s algorithm} was the first example of a quantum algorithm that could have significant real world implications by threatening the RSA cryptosystem \cite{RSA78}, which is  widely used to secure the Web. Its discovery initiated a flurry of research into quantum algorithms, a now burgeoning subfield of quantum information science (QIS), that has continued to the present. 
More recently, in 2019, experimental evidence of the increased computational power of quantum computers was provided via the first successful quantum primacy%
\footnote{We use the term \textit{quantum primacy} \cite{DGW21} in lieu of the original \textit{quantum supremacy} \cite{Pres12}, given recent discussions around the ethics of language in the sciences. In the context of this work, quantum primacy is considered a special case quantum advantage.} experiment on a 53 qubit superconducting device \cite{AAB+19}, with similar experiments following shortly after \cite{ZWD+20, WBC+21, ZCC+21}. While it has been debated whether these experiments represent true examples of quantum primacy%
\footnote{While each experiment represents a state-of-the-art demonstration of the capabilities of quantum devices, they are not without limitations. For example, whether the classical simulation performed in \cite{AAB+19} constitutes the practical limit of classical computation has been contested \cite{PGN+19,PCZ21}, although recent work has shown substantial advantages in energy consumption for the specific experiment \cite{VLB+20} (and in general\cite{BDF+21}). Further, while the superconducting devices in \cite{AAB+19, WBC+21, ZCC+21} are programmable, the photonic device in \cite{ZWD+20} was i) hard-coded to perform the specific task and ii) a Gaussian boson sampler (GBS) implementing a limited, non-universal model of quantum computation.}, they have nonetheless galvanized the QIS field around the practical potential of quantum computers in the near term. 

Quantum hardware has now entered the noisy intermediate-scale quantum (NISQ) era \cite{Pres18}, a stage of maturity characterized by devices with low qubit counts, high error rates, and correspondingly short coherence times. While it is unclear how long the transition from the NISQ era towards fault tolerant quantum computation (FTQC) will take – whether it occur by error correction \cite{AB08} or inherently fault tolerant hardware based on topological properties \cite{Kita03} – common estimates range from several years to several decades. Yet, with first generation NISQ devices moving from the lab to the cloud, now is an opportune time for computationalists in biology and medicine to begin exploring the value that quantum approaches may bring to their research toolbox.

\subsection{Biology and medicine as computational sciences}

Over the past three decades, biology and medicine have evolved into highly quantitative fields \cite{Mark17}. Areas of inquiry span from foundational questions on the origins of life \cite{HL15_b} and the relationships between protein structure and biological function \cite{BS01} to ones with a direct impact on clinical practice, such as those concerned with the oncogenesis of cancer \cite{HW00, HW11}, the development of novel drugs \cite{Jorg04}, and the precise targeting of therapeutics on the basis of genetic mutations \cite{DGO+06} and other clinical indicators \cite{CV15}. However, despite the substantial progress facilitated by computational methods and the expansion of high-performance computing (HPC) environments, fundamental constraints when modeling biological and clinical systems persist.

System complexity is one example. This constraint arises from both first-order biological complexity, as can be seen in the metabolic processes of individual cells \cite{NE21} or the binding of protein receptors to ligands \cite{WYZ21}, and higher-order clinical complexity, occurring at the intersection of complex biological, behavioral, socioeconomic, cultural, and environmental factors \cite{CFL18}. On one hand, this system complexity has made biological and clinical research a verdant playground for the development of many novel, efficient computational algorithms and approaches. On the other hand, practical algorithms typically manage system complexity via reductionist frameworks. A consequence of this is that existing computational models often fail to capture and reconcile important system dynamics. Quantum computers, if sufficiently robust ones can be built, promise to fundamentally reduce the algorithmic complexity of constructing and analyzing many of these models. This may allow solutions to many difficult problems to be computed with far greater efficiency, which could in turn reduce compute times and improve the fidelity of practical models.

The second constraint is one of scale. Looking to healthcare alone, as much as 153 exabytes of data were generated in 2013 with a projected annual growth rate of 48\% \cite{SHTR17}. Extrapolating this growth rate, it is plausible that over 2,300 exabytes were generated in 2020. Similar data challenges also exist in biology. For example, the high-throughput sequencing revolution has led to exabytes of highly complex genomic, epigenomic, transcriptomic, proteomic, and metabolomic data types (among others). To manage these large data volumes, centralized data repositories have proliferated (\textit{e.g.} see the ever-growing dbGaP \cite{MFJ+07} or the more recent Genome Data Commons \cite{HFA+21}). These massive data resources are crucial to the re-use of high-value data in secondary analyses and reproducibility studies. However, even with the wide use of HPC infrastructures, large bioinformatics and computational biology workflows often extend for days, weeks, or longer. In recent years, this challenge has grown with the expansion of other areas demanding significant computational resources. Examples include high-resolution imaging (\textit{e.g.} cryo-EM) and massive deep learning inference pipelines with on the order of $10^9$ (or greater) model parameters. While it is not anticipated that scalability constraints will be addressed by quantum computing technologies in the near term, FTQC devices may offer a partial solution to some of these challenges over the long term. 

\subsection{Approach}
\label{sec:Section-I-approach}

Given these challenges across biology and medicine and the potential of quantum computing, a common question among domain computationalists interested in quantum computing is: ``When will I be able to leverage quantum computing for [insert preferred application]?’' The answer to this question is complex and can be factored into a number of considerations:

\begin{itemize}[leftmargin=20pt]
    \item How can it be ascertained whether a problem will benefit from a quantum advantage?
    \item What scale of problem instance is required to meaningfully demonstrate such an advantage?
    \item What hardware and software are required to translate a quantum advantage from theory to practice?
    \item How can we detect and measure a practical quantum advantage once it has been achieved?
\end{itemize}

The goal of the work presented here is two-fold. The first goal is to consider what is known of the answers to these questions for a broad variety of potential applications in biology and medicine. To do this, we have sought to i) distill current knowledge around quantum advantages, quantum algorithms, quantum hardware, and a broad number of specific application areas in biology and medicine and ii) identify gaps in existing theory and implementations. Of course, we cannot hope to account for every possible application in such a large domain. Further, prior work \cite{HK10, CRA18, EWA+19, OSS+20, FG21} in this field has already considered aspects of these questions for a variety of application areas (\textit{e.g.} genomics, drug-design, clinical phenotyping, neurimaging). Thus, our second goal is to leverage our own exploration of these questions towards a greater framework to aid domain computationalists as they embark on their own explorations of quantum applications that have the potential to deliver practical quantum advantages in the near term.

\subsection{Impact}
\label{sec:Section-I-impact}

It is our hope that this work can serve as both i) a resource for computationalists in biology and medicine interested in developing quantum computing proof of principles for their field in the NISQ era and ii) a guide for quantum algorithms researchers interested in developing algorithms targeting applications in biology and medicine. As such, we have written this perspective with an interdisciplinary audience in mind and endeavored to minimize the need for a formal background in biology, medicine, quantum information, quantum algorithms, quantum hardware, and computational complexity. If additional background on the QIS field is desired by the reader, we refer them to the references highlighted in this footnote%
\footnote{See the following reviews and resources for more information on quantum information theory \cite{Hard01}, quantum algorithms \cite{Mont16, Jord21}, and quantum hardware \cite{Divi00}. For an exhaustive discussion of near term quantum algorithms and their prospects, we direct the reader to the recently published paper by Bharti et al. \cite{BCK+21}, alongside the more focused reviews on quantum machine learning \cite{BWP+17} and variational quantum algorithms \cite{CAB+20}. See the recent review by Eisert et al. for a discussion of hardware and software certification and benchmarking \cite{EHW+19} and a review by Gheorghiu et al. for a discussion of the more stringent verification of a quantum computation \cite{GKK19}. Finally, while the standard textbook for the QIS field is Nielsen and Chuang \cite{NC00}, see \cite{CEP+18} for a recent, application-oriented review of quantum algorithms and their circuit implementations.}. In addition, background on some of the topics in biology and medicine we discuss can be found here\footnote{See the following reviews and resources for more information on computational approaches in drug discovery \cite{Jorg04}, phylogenetics \cite{KYT20}, medical image segmentation \cite{HJH+19}, \textit{de novo} assembly \cite{SN16}, biological sequence error correction \cite{LBM16}, and deep learning applications in biology and medicine \cite{CHB+18}.}.

\section{The landscape of quantum advantages}
\label{sec:Section-II}

\textbf{The Landscape of Quantum Advantages} is a set of concepts that together are central to the identification, characterization, and realization of quantum advantages. These concepts, which we present below, include a classification scheme for quantum advantages, known hardware constraints that can influence their practical implementation, and context-based evidence levels for establishing their existence and practical realization.

\subsection{Theoretical quantum advantages}
\label{sec:Section-II-dtqs}

How can we define a quantum advantage from the theoretical perspective? While multiple mathematical formulations have been described in the literature (\textit{e.g.} see \cite{RWJ+14, BMG+20}), for our purposes, we simply state that a theoretical quantum advantage is defined by four key properties:

\begin{enumerate}[leftmargin=20pt]
    \item \textbf{Problem}: A formal computational problem.
    \item \textbf{Algorithms}: A classical algorithm and a quantum algorithm, each of which solve the computational problem.
    \item \textbf{Resources}: One or more resources, such as time, space, or data, that are consumed by both the classical and quantum algorithms.
    \item \textbf{Bounds}: Analytical bounds on the resource consumption (\textit{e.g.} a worst-case time complexity bound) for both the classical and quantum algorithms.
\end{enumerate}

A range of theoretical quantum advantages have been identified. For example, for the general problem of unstructured search, Grover’s algorithm \cite{Grov96, BBB+97} yields a quadratic advantage with a query complexity of $O(\sqrt{N})$ relative to the $O(N)$ queries required classically%
\footnote{Note that it is convention to take $N$ to mean $N=2^n$. This is due to the exponential state space (known as a Hilbert space) generated by an $n$-qubit quantum superposition.} (note that a query in this context can be thought of as a function call). On the other end of the spectrum, the \textit{k}-Forrelation problem \cite{BS20, SSW20} admits a quantum algorithm with the largest known (superpolynomial) quantum advantage – where ${\tilde{\Omega}(N^{1-1/k})}$ queries are required by a classical randomized algorithm (omitting logarithmic factors), only $\lceil k/2\rceil$ queries are required by the quantum algorithm. In most cases, theoretical quantum advantages should be thought of as loose approximations of the degree of quantum advantage that may be possible in practice.

\subsection{Classifying theoretical advantages}
\label{sec:Section-II-cts}

\begin{figure*}
    \centering
    \includegraphics[width=\textwidth]{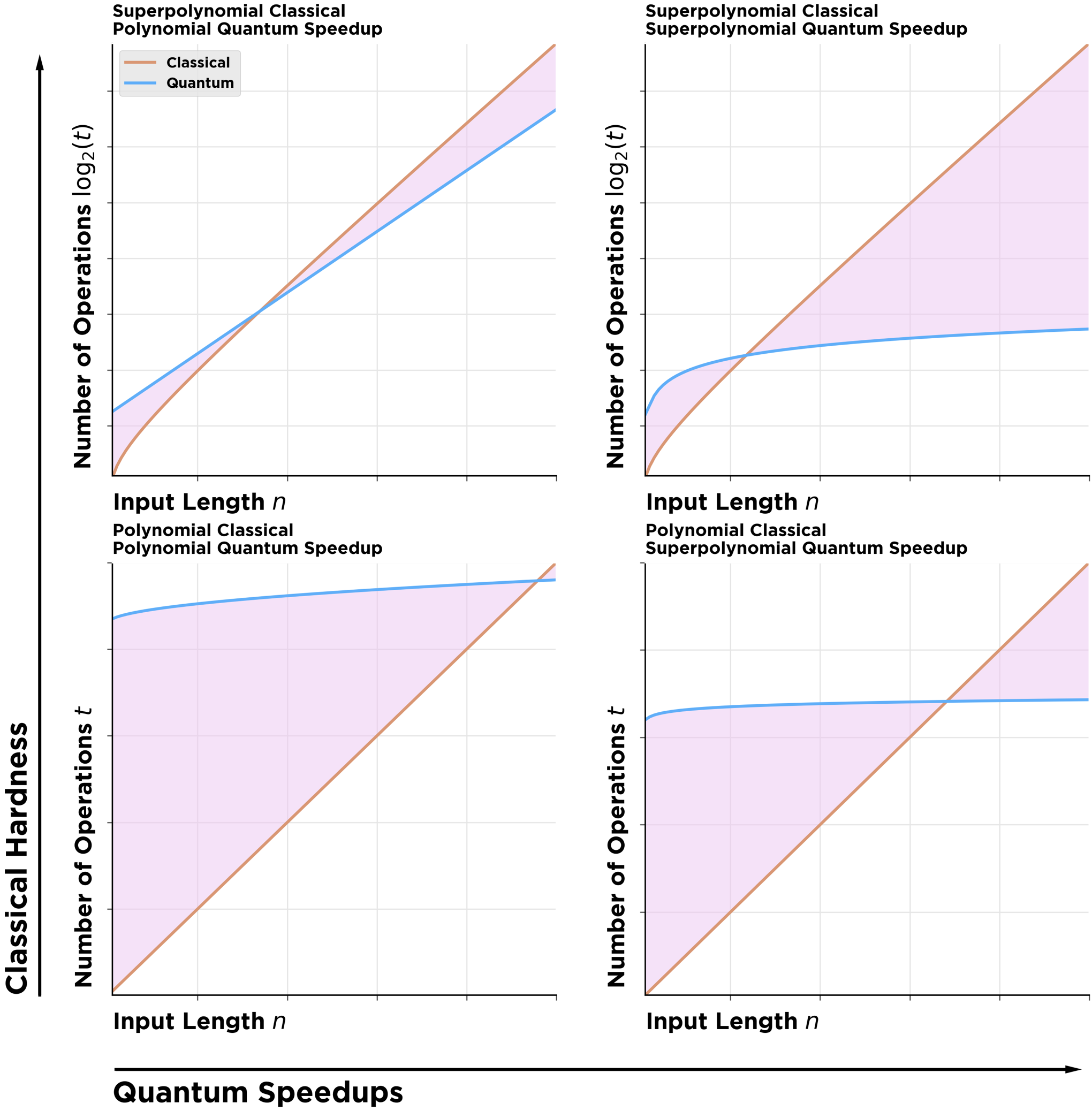}
    \caption{\textbf{Classifying quantum advantages.} A conceptual illustration of the classes of quantum advantages. We discuss four classes that we define across two axes: i) The classical resource consumption (vertical) and ii) the strength of the advantage (horizontal). Quantum advantages are expected to have variable computational overheads; as such, the pink region indicates where the quantum advantage begins. \textbf{Top Left:} Polynomial advantages on classically hard problems. Amplitude amplification and estimation \cite{BHM+02} provides a general framework to achieve polynomial quantum advantages on search and optimization problems. These techniques have led to quantum versions of dynamic programming algorithms \cite{ABI+18, Rona19}, which are relevant to many tasks in genomics, such as sequence alignment. \textbf{Top Right:} Superpolynomial advantages on classically hard problems. Hamiltonian simulation of strongly correlated fermionic systems is classically hard. Many quantum Hamiltonian simulation algorithms yield superpolynomial advantages \cite{CMN+18} and may lead to substantial improvements in drug discovery pipelines \cite{CRA18}. \textbf{Bottom right:} Superpolynomial advantages on classically easy problems. Matrix inversion is among a set of linear algebra subroutines that are central to machine learning. In particular, inverting a feature covariance matrix is common to many classical machine learning approaches, such as support vector machines and Gaussian processes. The quantum linear systems algorithm (QLSA) \cite{HHL08} performs this crucial subroutine for a variety of QML algorithms (\textit{e.g.} see \cite{RML14, KP17}) and runs in polylogarithmic time relative to the matrix dimensions. While this implies a superpolynomial advantage (omitting the complexity due to the condition number), practical realizations are expected to depend on matrix rank and may require the development of quantum random access memory (QRAM) \cite{GLM08a, GLM08b, AGJ15, HZZ+19, CDE+21}. \textbf{Bottom left:} Polynomial advantages on classically easy problems. Grover’s algorithm \cite{Grov96} for unstructured search is the classic example. Quantum algorithms in this class will likely be used as subroutines in conjunction with other quantum algorithms that offer greater advantages.}
    \label{fig:fig1}
\end{figure*}

A theoretical quantum advantage can be classified on the basis of two factors (\textbf{Figure \ref{fig:fig1}}). The first one relates to the classical hardness of the computational problem, which is defined by the best known classical algorithm%
\footnote{In practice, we are often interested in the best \textit{available} classical algorithm, which may be identified by a mixed criteria. Examples of this criteria include i) the most commonly used practical algorithm for a given application, ii) the most advanced algorithm implemented for that application, or iii) the algorithm for which the underlying operational assumptions best comport with the ones required for the comparison being made.} or a provable upper bound. In particular, a computational problem may be classified as \textit{easy} or \textit{hard} according to whether its classical algorithmic complexity (typically for a worst case input) is polynomial or superpolynomial, respectively \cite{Cobh65}. The second factor relates to the size of the advantage yielded by the quantum algorithm relative to the classical algorithm. Often, these advantages result from a reduction in complexity for a computational resource. Like algorithmic complexity, the size of an advantage advantage may also be classified as polynomial or superpolynomial. Here, we would like to make clear that the terms ``easy'' and ``hard'' refer only to classical computability; they are not intended to refer to the attainability of a practical quantum advantage or the difficulty of implementing a quantum approach. 

For the rest of this subsection, we discuss paradigmatic examples of each class of quantum advantage. While we briefly highlight some relevant algorithms and applications in biology and medicine, we provide a much more detailed account of potential applications in \textbf{Section \ref{sec:Section-IV}}.

\textbf{Polynomial advantages for easy problems.} The most well-known algorithm in this class is Grover's algorithm for unstructured search, which has a worst-case complexity of $O(\sqrt{N})$ relative to the $O(N)$ complexity of its classical counterpart \cite{Grov96}. At its core, Grover's algorithm uses amplitude amplification \cite{BHM+02}, a general quantum algorithm technique that yields polynomial advantages without requiring a specific problem structure. Many other quantum algorithms also fall under this class of quantum advantage. These include ones for convex optimization \cite{CCL+18, VGG+19}, semi-definite programming \cite{BS17}, and calculating graph properties (\textit{e.g.} bipartiteness, st-connectivity, and cliques) \cite{MSS07, ACL11, CMB16}. From the perspective of biology and medicine, some problems in network and systems biology can be cast as convex optimization or network inference problems incorporating graph property testing%
\footnote{Recent work suggests that, given certain reasonable assumptions, graph property testing algorithms will only be able to yield polynomial quantum advantages \cite{ACL11, CW20, BP20}.}. Some practical examples of these problems include inferring the characteristics and behaviors of gene regulatory, protein interaction, and metabolic networks. 

\textbf{Polynomial advantages for hard problems.} These include many $\textrm{NP}$-hard optimization problems that can also benefit from amplitude amplification. Examples of quantum algorithms in this class include ones for constraint satisfaction \cite{Mont18} and combinatorial optimization \cite{ABI+18, Rona19, Mont19}. Notably, algorithms in this class were among the first to target applications specific to biology and medicine, including sequence alignment \cite{Holl00, LMB06, PK19} and the inference of phylogenetic trees \cite{EJ11}. Sequence alignment, in particular, represents a crucial computational primitive for many tasks in bioinformatics and computational biology. 

\textbf{Superpolynomial advantages for easy problems.} While no algorithms specific to problems in biology or medicine are known to exhibit this class of advantage, a number of quantum machine learning (QML) algorithms with general relevance do. Perhaps the most prominent example is the quantum linear systems algorithm (QLSA) \cite{HHL08} for high-rank matrices, which led to the development of many early QML algorithms. Another example is a quantum algorithm for pattern matching \cite{Mont17}, which yields a superpolynomial advantage as the size of the pattern approaches the size of the input. Quantum algorithms in this class typically exploit compact quantum data encodings, such as amplitude encoding. These encodings allow for certain computations to be performed using a polylogarithmic number of qubits and operations relative to their comparable classical algorithms. However, they may also be subject to several practical constraints related to data input and output sampling \cite{Aaro15}.

\textbf{Superpolynomial advantages for hard problems.} The most widely known example of an algorithm in this class may be Shor’s algorithm for factoring integers \cite{Shor97}, a problem that has no direct relevance to biology or medicine. Quantum algorithms for simulating quantum physics \cite{AL97, KJL+08, Chil10, BC12, BCC+15, LC17}, on the other hand, may find significant applications. In principle, these algorithms could provide superpolynomial advantages for a vast range of hard problems related to simulating physical systems%
\footnote{Indeed, quantum simulation is viewed by many as the ``killer app'' for quantum computing in a wide variety of fields, including chemistry, material science, and molecular biology.}. Examples include characterizing the ground states of biologically relevant small molecules, the behavior of chemical solutions, and the ternary and quaternary structures of biological molecules \cite{KA09}. In general, superpolynomial advantages for hard problems are the most desirable class of quantum advantage.

\subsection{Context-based evidence for quantum advantages}
\label{sec:Section-II-elqs}

A quantum advantage may be supported by multiple forms of evidence that can vary by its context. At the highest level, these contexts can be theoretical or empirical. In the theoretical context, abstract evidence is collected to answer whether a quantum algorithm can yield an advantage over the most efficient classical algorithm. In contrast, empirical evidence is collected to answer whether a specific quantum algorithm and device can, together, yield an observable advantage over a classical approach given a well-defined problem and context-dependent metric. Importantly, i) these contexts are often not mutually exclusive (\textit{i.e.} a degree of overlap may exist) and ii) while an experimental advantage may also be practical, evidence from an operational context is best suited to addressing questions around the practical value of an observed quantum advantage. In this section, we describe these contexts in detail and their implications for the evidence required to establish a quantum advantage. 

\textbf{Theoretical advantages} result from an improvement in analytical bounds on resource efficiency by a quantum algorithm relative to a well-motivated classical counterpart. The core resource in question typically involves units of time, space, or information (\textit{e.g.} samples from an experiment). In practice, these units of comparison may be gates, queries, bits, or error rates. Theoretical advantages are defined mathematically, may be conjectured, and are often contingent on well-founded assumptions from computational complexity theory. Crucially, they need not be application-oriented%
\footnote{While a theoretical algorithm may indicate that a quantum advantage exists, the theoretical algorithm need not be implementable or relevant to a known practical problem.}. Instead, they may be conceptual – designed with the express intent of demonstrating that an advantage exists for an artificial task or broader class of problems that have theoretical interest. Examples of conceptual quantum algorithms yielding theoretical advantages include the Deutsch-Jozsa algorithm \cite{DJ92}, Bernstein-Vazirani algorithm \cite{BV97}, Simon’s algorithm \cite{Simo93}, and one for solving the Forrelation problem \cite{AA14b}. Despite their theoretical motivation, practical applications may nonetheless follow from conceptual algorithms. One example of this can be seen with Shor's algorithm for prime factorization \cite{Shor94}, which was inspired by Simon's algorithm \cite{Simo93}.

\textbf{Experimental quantum advantages} are a type of empirical quantum advantage observed in an experimental context. Crucially, they require the comparative analysis of computational metrics that precisely measure the advantage in question%
\footnote{See \cite{BY19} for a review of and general framework for benchmarking metrics of a computation.}. To facilitate these comparisons, these metrics are ideally defined over both quantum and classical approaches. While theoretical evidence (\textit{i.e.} proofs, conjectures, and numerical models) may support an experimental advantage \textit{a priori}, it is the computational benchmarking of an implementation that serves as the evidence for establishing an experimental quantum advantage. For instance, the first quantum primacy experiment \cite{AAB+19} represented a real-world demonstration of an experimental quantum advantage. First, a theoretical advantage was proven and numerically estimated \cite{BIS+18a}, then initial experimental work soon followed, which yielded the first hint of such an experimental advantage \cite{NRK+18}. Finally, this culminated in an experimental demonstration of the quantum advantage – as evidenced by substantial cross-entropy benchmarking \cite{AAB+19}. Similar trajectories are observable for other quantum primacy experiments, such as one recently demonstrated on a non-universal photonics-based device \cite{ZWD+20, AA13, WQD+19}.

\textbf{Operational quantum advantages} result from the successful translation of an experimental advantage to an applied setting. As such, in addition to the computational benchmarking required to validate the experimental advantage, this type of advantage may require the definition and measurement of an operational metric – a key performance indicator (KPI) – to gauge the extent of the practical advantage when deployed. Accordingly, this type of advantage considers the greater context outside of the computational environment. In particular, it can be viewed as one that additionally integrates the organizational, economic, and social context of the target application. Challenges to realizing an operational quantum advantage may include organizational inertia, entrenched support for incumbent classical methods, and difficulties in the integration of the experimental quantum advantage into existing software, hardware, and network infrastructure. To date, no obvious demonstration of an operational quantum advantage has occurred and estimates of when such an advantage may occur vary greatly.

\textbf{From theory to practice.} The context of a quantum advantage is used to determine what evidence is required to establish it. While theoretical evidence has historically preceded experimental evidence, the chronological order of evidence levels may vary. This is expected to become increasingly apparent as interest in practical quantum approaches and access to near term quantum hardware expands. In the absence of supporting theoretical evidence, the pursuit of experimental advantages may be motivated instead by expert intuition (\textit{e.g.} around the structure of a computational problem and how quantum information may be beneficial to finding its solution). While such an intuition-based approach provides practical flexibility, it should be cautioned that it may, too, be a source of bias \cite{BAF21}. To manage this risk for specific applications, robust benchmarking and the publication of full experimental evidence (e.g. data and analysis code) through open science tools will be key to community verification of claimed empirical quantum advantages over the near term, especially in operational contexts. 

\subsection{Practical constraints on theoretical advantages}
\label{sec:Section-II-pcts}

Translating a theoretical advantage into an experimental one is often challenging. These challenges largely arise from the practical constraints of NISQ hardware. In this section, we discuss some of these constraints and how they inform the translation of theory into practice. For an overview of the potential feasibility of a variety of quantum advantages with respect to quantum hardware, see \textbf{Table \ref{table:table1}}.

\begin{table*}[!ht]
    \centering
    \begin{adjustwidth}{-0.5in}{0.5in}
        \begin{tabular}{
            >{\raggedright\arraybackslash}p{.12\textwidth} 
            >{\raggedright\arraybackslash}p{.12\textwidth} 
            >{\raggedright\arraybackslash}p{.13\textwidth} 
            >{\raggedright\arraybackslash}p{.13\textwidth} 
            >{\raggedright\arraybackslash}p{.12\textwidth} 
            >{\raggedright\arraybackslash}p{.12\textwidth} 
            >{\raggedright\arraybackslash}p{.12\textwidth} 
            >{\raggedright\arraybackslash}p{.07\textwidth} 
            >{\raggedright\arraybackslash}p{.07\textwidth}}
            \toprule \\
            \textbf{Quantum Advantage} &
            \textbf{Resource} &
            \textbf{Example Application} &
            \textbf{Quantum Algorithm(s)} & 
            \textbf{Quantum Complexity} &
            \textbf{Classical Complexity} & 
            \textbf{Classical Constraint} &
            \textbf{NISQ} & 
            \textbf{FTQC} \\ 
            \\ \midrule \\
                Super-\newline polynomial &
                Operations &
                Hamiltonian Simulation &
                \cite{AL97, BCC+15, LC17, Lloy96, LC19,  SMK+20} &
                Polynomial &
                Super-\newline polynomial &
                Complexity &
                Likely &
                Yes \\
            \\ \vspace{0pt} \\
                Super-\newline polynomial &
                Operations & 
                Matrix Inversion &
                \cite{HHL08, Amba12, CJS13, CKS17, WZP18} &
                Poly-\newline logarithmic &
                Polynomial &
                Very large data &
                Possibly &
                Likely \\
            \\ \vspace{0pt} \\
                Polynomial &
                Queries &
                Unstructured Search &
                \cite{Grov96, BHM+02} &
                Polynomial &
                Polynomial &
                Very large data &
                Unlikely &
                Possibly \\
            \\ \vspace{0pt} \\
                Polynomial &
                Queries &
                Dynamic Programming &
                \cite{ABI+18, Rona19} &
                Super-\newline polynomial &
                Super-\newline polynomial &
                Complexity &
                Unlikely &
                Possibly \\
            \\ \vspace{0pt} \\
                Polynomial &
                Samples &
                Machine Learning &
                - &
                Polynomial &
                Polynomial &
                Sample generation &
                Possibly &
                Possibly \\
            \\ \vspace{0pt} \\
            Super-\newline polynomial &
                Samples &
                Machine Learning &
                - &
                Polynomial &
                Super-\newline polynomial &
                Complexity; quantum data &
                Unlikely &
                Possibly \\
            \\ \bottomrule
        \end{tabular}
    \end{adjustwidth}
    \vspace{12pt}
    \caption{\textbf{Quantum advantages and their expected feasibility by hardware.} The type of advantage yielded by a quantum algorithm informs the degree of hardware maturity that is expected to be required for its practical realization. Whereas superpolynomial quantum advantages on classically hard problems may be readily attained by NISQ devices in the coming years, other forms of advantage (\textit{i.e.} superpolynomial advantages on classically easy problems and polynomial advantages on both classically easy and hard problems) will likely require greater hardware maturity, up to FTQC. In the right-most columns, we indicate the hardware paradigm that is expected to be required for the stated quantum advantage to be realized by a practical application. Note that sample complexity advantages are currently supported by theoretical proofs; algorithms with the potential to demonstrate them in practice are under investigation.}
    \label{table:table1}
\end{table*}

\textbf{Logical versus noisy qubits.} Theoretical quantum algorithms tend to be modeled with FTQC devices – ones implementing logical qubits upon which error-free gates, state-preparation, and measurement are performed. Current hardware remains far from such a scalable device. Instead, existing NISQ devices have dozens to hundreds of error-prone qubits. The error characteristics of quantum devices arise from a number of error types. These include state preparation and measurement (SPAM) error, gate (operation) errors, emergent errors, such as cross-talk \cite{SPR+19, EWP+19, MMM+20}, and systematic errors due to device calibration or fabrication defects. While these errors can in principle be factored into their constituent phase and bit-flip components, the heterogeneity of their generating processes will likely require an all of the above approach to realize scalable FTQCs with coherence times sufficiently long to run large quantum circuits with polynomial depth.

Approaches to realizing FTQC are expected to require error mitigation techniques\footnote{Notably, recent work has shown that error mitigation must be implemented with care as it can decrease the effectiveness of circuit parameter learning in some cases \cite{WCA+21}} (\textit{e.g.} see \cite{WCA+21, TBG17, LB17, MYB19, SBS+19, BW20, AAB+20_a, BGC+21, BGK+21, TEM+21, NKS+21}) to bring hardware error rates below a fault-tolerance threshold \cite{AB08}. Below this threshold, it may be possible to use error correcting codes (\textit{e.g.} see the CSS code \cite{Shor95, CS96}, surface code \cite{FSG12}, XZZX surface code \cite{ATB+20}, and honeycomb code \cite{HH21}) to enable practically unlimited coherence times%
\footnote{The required fault-tolerance threshold depends on a variety of factors, including the hardware error characteristics, connectivity map, and the error correction code being used.}. 

Numerical simulations of error correction approaches imply that the overheads incurred by some of these strategies will place limits on the class of quantum advantages that can be achieved for many years%
\footnote{A recent case study \cite{GE19} found that high constant time overheads represent a significant challenge in implementing Shor’s algorithm on a benchmark application in cryptography. As modeled, this overhead is due to magic state distillation \cite{BK05, FMM+12}, a process required for error correcting non-Clifford gates (crucial operations for universal quantum computation). While the overhead of magic state distillation has long been viewed as necessary to FTQC, it is notable that this has recently been called into question (\textit{e.g.} see \cite{Brow20}).}. Indeed, some polynomial advantages \cite{BMG+20} and even superpolynomial advantages \cite{GE19} using the surface code may require very large numbers of qubits to encode the necessary logical qubits. Fortunately, the rapid pace of progress and many avenues being explored give much reason for optimism. One recent example lies in a numerical simulation of the honeycomb code \cite{HH21}. In particular, this numerical simulation indicated that achieving computations on circuits with trillions of logical operations may be possible with as few as 600 qubits\cite{GNF+21}, given modest assumptions around qubit error rates.

\textbf{Qubit connectivity.} With few exceptions, NISQ computers are expected to have limited qubit connectivity. In contrast, the mathematical proofs behind theoretical quantum advantages often assume all-to-all connectivity within qubit registers. Fortunately, a growing body of evidence \cite{Brie16, Herb20} around the number of SWAP gates \cite{BBG+13, SGH20} required to emulate all-to-all connectivity on lower connectivity architectures indicates this may be a soft constraint on practical quantum algorithms. For example, one recent numerical simulation demonstrated that all-to-all connectivity can be simulated with a 2D grid architecture (with logarithmic or sublinear bounds on the overhead) for three quantum subroutines central to quantum simulation, machine learning, and optimization algorithms \cite{HJG+20}. This work also highlighted that logarithmic bounds may also be realized in architectures with even greater connectivity constraints provided sufficient additional qubits are available%
\footnote{These numerical simulations of quantum circuits included ones for the quantum Fourier transform, Jordan-Wigner transform, and Grover diffusion operator, each compiled using three connectivity maps representative of existing device architectures: All-to-all with $n(n-1)/2$ edges, a 2D square grid with $\sqrt{n} \times \sqrt{n}$ dimensionality and $2(n-\sqrt{n})$ edges, a ladder with $n/2 \times 2$ dimensionality and $3n/2-2$ edges, and a linear nearest neighbor graph with degree-1 terminal qubits and $n-1$ edges.}. Altogether, these numerical simulations imply that qubit connectivity on its own may present only a mild constraint when pursing practical quantum advantages.

\textbf{Input constraints.} Theoretical quantum algorithms typically use a combination of abstract input models (\textbf{Table \ref{table:table2}}). These input models can either be oracles – ``black-box'' functions for which the form of the input and output is known but the implementation of the function is not – or data inputs. Both oracles and data inputs can be either quantum or classical. Practically, oracles can be viewed as algorithm subroutines or queryable data structures. Importantly, the type of oracle (\textit{i.e.} quantum or classical) can either improve or limit the feasibility of implementing a quantum algorithm on a NISQ device. Whereas quantum oracles may place additional complexity requirements with respect to the number of qubits or gates required to implement a quantum circuit, classical oracles may mitigate the size of a quantum circuit by offloading classically efficient computations to a classical device. Similarly, the input of data typically requires either the encoding of classical information into qubits (\textbf{Box} \hyperref[box:box1]{\textbf{1}}) or the loading of a quantum state. These data input steps generally demand additional circuit depth. 

When theoretically analyzing quantum algorithms, the complexity overheads of oracles and data input are sometimes omitted from the stated complexity. Despite these omissions, these subroutines can strongly influence whether an empirical quantum advantage is possible. One prominent example exists for the class of superpolynomial advantages on classically easy problems%
\footnote{Quantum algorithms yielding this class of advantage often leverage amplitude, or another similarly efficient encoding (\textit{e.g.} see \cite{APP+20}), requiring on the order of $\log_2 N$ qubits.}. In principle, for this class, the data input must be performed in $O(\mathrm{polylog}(n))$ to maintain the quantum advantage, a point highlighted in a discussion \cite{Aaro15} of the potential limitations of the QLSA%
\footnote{In particular, the QLSA embeds the data matrix into the amplitudes of the quantum state, which implies that a quantum state with amplitudes proportional to the values in the data matrix must be prepared in $O(\mathrm{polylog}(n))$ time \cite{Aaro15}. To do this, the QLSA assumes access to a quantum random access memory (QRAM) \cite{GLM08a, GLM08b, CDE+21}, a hardware device expected to require fault tolerance. Additionally, given the efficiency of the amplitude encoding used for the data input, the size of the measured output is limited proportionally (\textit{i.e.} $\log_2 N$) due to Holevo's bound \cite{Hole73}. To sample the full solution vector, a polynomial number of samples (and an i.i.d. – independent and identically distributed – assumption over output of the quantum circuit) is required, abrogating the superpolynomial speedup.}\footnote{It is notable that recent work has indicated that this particular input constraint on the QLSA may be less severe in practice, given a modest assumption \cite{ZFR+19}.}. 
\newline
\begin{mdframed}
    \textbf{Box 1: Data encoding methods.} While many data encoding methods for quantum algorithms exist, three are commonly used:
    \begin{itemize}[leftmargin=*]
        \item\textbf{Computational basis encoding} allows for an $n$-bit binary strings to be encoded into $n$ qubit basis states. It can be thought of as a quantum version of a classical computer writing binary data to memory.
        \item\textbf{Amplitude encoding} allows for the input of real values into the $N=2^n$ amplitudes (\textit{i.e.} relative phase) of a quantum system using $\log N$ qubits. 
        \item\textbf{Angle encoding} allows for the efficient encoding of $n$ real-valued inputs by parameterizing $n$ qubits with the input values and then computing their tensor product.
    \end{itemize}
    For hybrid quantum-classical algorithms (discussed in \textbf{Section \ref{sec:Section-III}}), the type of data encoding method used can significantly influence both the expressivity of the parameterized quantum circuit (PQC) \cite{LC20, SSM21} and the appropriate procedure for sampling the output distribution of the quantum circuit. 
    \label{box:box1}
\end{mdframed}

\begin{table*}[!h]
    \centering
    \begin{adjustwidth}{0in}{0in}
        \begin{tabularx}{\textwidth}{
            >{\raggedright\arraybackslash}p{.12\textwidth}
            >{\raggedright\arraybackslash}p{.25\textwidth}
            >{\raggedright\arraybackslash}p{.08\textwidth}
            >{\raggedright\arraybackslash}p{.08\textwidth}
            >{\raggedright\arraybackslash}p{.38\textwidth}}
            \toprule \\
                \textbf{Input Model} & 
                \textbf{Definition} & 
                \textbf{Quantum} & 
                \textbf{Classical} & 
                \textbf{Examples}  \\
            \\ \midrule \\
                Standard & $x=(x_1,...,x_n) \in \{0,1\}^n$ & Yes & Yes & Factoring and other number theory problems (\textit{e.g.} Shor’s algorithm), 3-SAT and combinatorial optimization, variational quantum eigensolver (VQE). \\
            \vspace{0pt} \\
                Oracle & $O\vert i, a \rangle = \vert i,a + x_i \rangle$ & Yes & Yes & OR (Grover search), max, approximate counting, NAND tree, collision, graph property testing, hidden subgroup problem (HSP), welded trees. \\
            \vspace{0pt} \\
                Quantum Data & given state $\vert \psi \rangle$ & Yes & No & Quantum Fourier transform (QFT), state tomography, Hamiltonian simulation, quantum linear systems solver (HHL), learning with quantum examples. \\
            \vspace{0pt} \\
                Quantum \newline Oracle & the ability to implement a unitary $\mathit{U}$ and controlled unitary $\mathit{C_U}$ & Yes & No & Phase estimation, quantum sensing and process tomography, qubitization, and singular value transform. \\
            \vspace{0pt} \\
                Coresets \cite{Harr20} & $x=(x_1,...,x_n) \in \Sigma^n$ \newline and oracle access $f: \Sigma \times Y \mapsto \mathbb{R}$ & Yes & Yes & Potential applications identified include $k$-means clustering, logistic regression, zero-sum games, and boosting. \\
            \\ \bottomrule
        \end{tabularx}
    \end{adjustwidth}
    \vspace{12pt}
    \caption{\textbf{Input models for quantum algorithms.} A major challenge in quantum algorithm development is determining the most efficient way to input classical data and prepare quantum states. The standard, oracle, quantum data, and quantum oracle models represent four basic approaches (along with many variants) used in developing theoretical quantum algorithms. One such variant is the coreset input model, which can reduce the size of the input and be viewed as a hybrid of the standard and oracle models. Sometimes, a quantum algorithm will leverage multiple input models (\textit{e.g.} both quantum data and a quantum oracle). This table was reproduced (with minor modification) from \cite{Harr20}.}
  \label{table:table2}
\end{table*}

Many quantum algorithms for preparing initial quantum states already exist \cite{APP+20, MVB+05, SBM06, HM20, ZYY21}. Nonetheless, over the near term, low qubit counts and circuit depths are expected to limit the realization of many quantum advantages in practice. To mitigate this challenge, a variety of approaches have been proposed. These include the ``coresets'' input model \cite{Harr20}, a technique with conceptual similarities to importance sampling, for which one recent practical implementation has been demonstrated \cite{TGA+21}. Similarly, data transformations designed to compress the correlation structure of features into a lower dimensional space may also prove useful. Further, hybrid quantum-classical approaches (discussed in detail in \textbf{Section \ref{sec:Section-III}}) and recent work highlighting the relationship between data encodings and learned feature embeddings \cite{SSM21, SK19, LSI+21} have substantially improved our understanding of efficient data input for near term devices. In particular, hybrid quantum-classical techniques can allow for substantial reductions in circuit width and depth for many applications \cite{CAB+20}. Thus, while input constraints are expected to remain prevalent in the near term (particularly for algorithms requiring polynomial qubits and circuit depths), the substantial work on mitigating them and recent progress with hybrid approaches provides a growing toolkit for managing them.

\textbf{Output constraints} can largely be attributed to two highly-related factors arising from the physics of quantum information. The first is Holevo's bound \cite{Hole73}, a theoretical result that limits the amount of information that can be efficiently retrieved from $n$ qubits to $n$ classical bits. Naïvely, this implies that the size of a solution output by a quantum circuit must be (at most) polynomial in the number of qubits. The second constraint relates to the probabilistic nature of quantum information. In particular, even if a solution can be output within the space afforded by Holevo's bound, the number of quantum circuit evaluations (\textit{i.e.} samples) to identify it with confidence must also have polynomial scaling for efficient computability. 

Together, these constraints can limit the practicality of implementing certain classes of quantum advantage. One illustrative example relates to the previously mentioned QLSA. In particular, the number of samples required to output the complete solution vector is expected to scale polynomially with $n$. Given the algorithm runs in $O(\mathrm{polylog}(n))$ operations, fully extracting the solution vector abrogates the superpolynomial speedup \cite{Aaro15}. As such, this particular algorithm (and many related ones) are expected to deliver a superpolynomial advantage only for sampling statistics of the solution vector or when calculating some global property of the state. More generally, these output constraints emphasize the importance of choosing an appropriate input encoding in anticipation of the number of classical bits that will be required to output the solution.

\section{The pursuit of NISQ advantages}
\label{sec:Section-III}

\subsection{A brief overview of variational quantum algorithms}
\label{sec:Section-III-vqa}

\begin{figure*}[!ht]
    \centering
    \includegraphics[width=\textwidth]{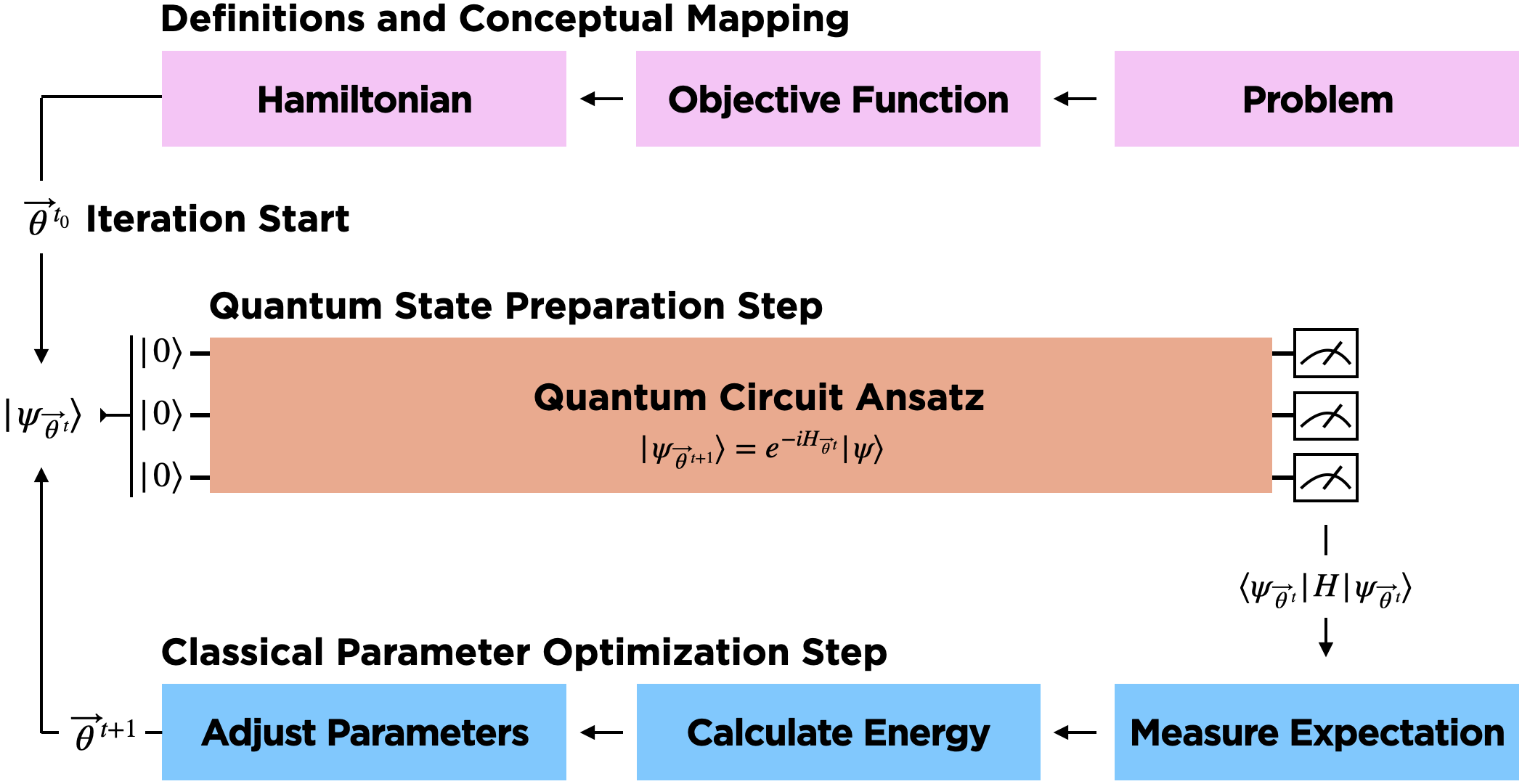}
    \caption{\textbf{Schematic of a VQA.} Development of a VQA begins with the mapping of a problem and objective function to a Hamiltonian operator, $H$. Following this mapping, execution of the VQA proceeds as follows: First, the quantum circuit is initialized to the state $\vert 0 \rangle ^ {\otimes n}$. The first execution of the parameterized quantum circuit can be viewed as the initial ansatz, $\vert \psi_{\vec{\theta}} \rangle $. Next, a measurement is performed to extract the information required to compute the loss of the objective function (\textit{e.g.} the expectation of the Hamiltonian operator, $\mathcal{L}(\vec{\theta})=\langle \psi_{\vec{\theta}} \vert H \vert \psi_{\vec{\theta}} \rangle$). Using this measurement, a classical optimizer then computes updates to the parameters to minimize the loss of the objective function (\textit{e.g.} a simple form of gradient descent uses the update rule $\vec{\theta}^{t+1} := \vec{\theta}^t - \eta \nabla \mathcal{L}(\vec{\theta}^t)$, where $\eta$ is a hyperparameter for the gradient step size or learning rate and $t$ is the time step). This proceeds iteratively until the system has converged from the initial ansatz to the lowest energy state of $\vert \psi_{\vec{\theta}} \rangle$. This low energy state represents both an approximation of and upper bound on the lowest energy state of the Hamiltonian operator.}
    \label{fig:fig2}
\end{figure*}

Understanding empirical quantum advantages and their computational benchmarking has taken on additional importance with the development of hybrid quantum-classical approaches. Many of these hybrid approaches developed for NISQ devices can be described as variational quantum algorithms (VQAs). In general, a VQA leverages three basic components: i) a parameterized quantum circuit (PQC), ii) an objective function, and iii) a classical optimizer\footnote{It is notable that, in addition to the many gradient-based optimizers seen in classical deep learning libraries \cite{Rude17}, VQAs may leverage quantum-specific optimizers. In particular, the parameter-shift rule \cite{LYP+17, MNK+18_a} represents a favored gradient-based approach developed expressly for VQAs. Recent work on this optimizer has led to a broadening of the gate sets to which it can be applied \cite{Croo19, BC21} and the use of adaptive shots \cite{KAC+20, ACS+20}, which leverage information from prior circuit evaluations to reduce the number of measurements required. Other optimizers include a quantum analog of natural gradient descent \cite{SIK+20, Yama19} and derivative-free methods \cite{MRB+16}.} (\textbf{Figure \ref{fig:fig2}}). Examples include the variational quantum eigensolver (VQE) \cite{PMS+14} and many variational QML algorithms. Crucially, they allow for substantial flexibility in their implementation and application. 

To design a VQA, the first step is to define a problem Hamiltonian $\mathit{H}$ (\textit{i.e.} a Hermitian operator that, in the context of physical simulation, corresponds to the kinetic and potential energy of the system), and associated objective function, $\mathcal{L}$. This is followed by the design of an appropriate PQC (sometimes called an ansatz) composed of both fixed and variational gates, which may be single-, two-, or multi-qubit gates. The circuit is controlled by parameters $\vec{\theta} \in \mathbb{R}^d$, which are made to vary by a classical optimizer, which seeks to minimize the the objective function. These parameters may control a variety of aspects of the quantum circuit, such as quantum gate rotation angles (\textit{e.g.} $R_y(\vec{\theta})$), whether a layer or unitary in the circuit should execute, or whether execution should halt upon a condition being reached. As an example, a common VQA approach to quantum simulation is to minimize the expectation value of the Hamiltonian $\mathcal{L}(\vec{\theta})=\langle \psi_{\vec{\theta}} \vert \mathit{H} \vert \psi_{\vec{\theta}} \rangle$ with respect to the parametric state $\ket{\psi_{\vec{\theta}}}$ prepared by the quantum circuit. Doing so leads to an approximation of the ground state or minimum eigenvalue of $\mathit{H}$, providing an upper bound on the ground state energy of the Hamiltonian.

Using this approach, VQAs may solve problems in the near term that would be otherwise infeasible due to the inherent coherence limitations of NISQ devices. Further, by leveraging classical post-processing techniques on the output observables \cite{OG18, CRD+18, KTC+19}, some theoretical evidence suggests that VQAs can be made resistant to errors.

\subsection{Variational quantum simulation}
\label{sec:Section-III-vqs}

Quantum simulation is one of the most anticipated applications for quantum computers, particularly in the domains of condensed matter physics, quantum chemistry, material science, and structural biology. Within these fields, a core goal is the scalable simulation of large, strongly correlated fermionic systems with an accuracy far superior to current approximate solutions from classical computation. Realizing this goal may precipitate a novel paradigm within these fields where our ability to understand the material world – superconductors, meta materials, chemistries, catalysts, proteins, and pharmaceutical compounds – is vastly improved. Given the many superpolynomial advantages offered by quantum simulation algorithms, it is highly plausible that variational quantum simulation (VQS) will yield many practical applications in the near to medium term. These applications may in turn have a significant impact across a broad variety of research fields and industrial sectors.

The first VQS algorithm (and VQA) was the variational quantum eigensolver (VQE) \cite{PMS+14}. The VQE was originally conceived to compute quantitative properties of molecular Hamiltonians. The first problem instance targeted by the VQE was finding the ground state of helium hydride \cite{PMS+14}. Since then, various implementations of the VQE approach have been used to simulate molecular binding \cite{DMH+18} and dissociation \cite{OBK+16, GEB+19}, dipole moments \cite{RGT+20}, chemical reactions \cite{AAB+20_b}, and other quantum mechanical systems and their quantities. So far, these practical implementations have all focused on simple molecules with relatively few atomic nuclei%
\footnote{Classical numerical simulations of VQEs and related approaches have been performed for larger systems; for one example, see \cite{CHZ+21}} – simulations that are classically tractable (and with greater accuracy). These molecules include multiple hydrides, such as ones of beryllium (BeH2) \cite{KMT+17} and sodium, rubidium, and potassium (NaH, RbH, and KH, respectively) \cite{MPJ+19}. While these represent promising early results, significant work remains to scale VQS approaches to larger molecules that have biological relevance, such as proteins, enzymes, nucleic acids, and metabolites. In fact, the simulation of such a biomolecule to high accuracy would likely represent a clear demonstration of an experimental quantum advantage.

A number of barriers currently prevent VQS advantages from being realized. One barrier is the number of measurements required by VQS algorithms to attain chemical accuracy competitive with existing classical simulation techniques \cite{WHT15, GRB+20}. Fortunately, substantial recent work suggests that this measurement problem may not be insurmountable \cite{IYL+19, VYI20, HMR+21, WKJ+21}. Another barrier relates to whether VQAs are sufficient to efficiently model the physical aspects of Hamiltonian simulation problems that challenge classical algorithms. Without this capability, the superpolynomial advantages established by existing theoretical quantum simulation algorithms \cite{AL97, KJL+08, KA09, Lloy96, AL99, ADL+05} may not be realized by VQS approaches, which are inherently heuristic. 

It seems unlikely that this will be the case. Significant theoretical \cite{MRB+16, YEZ+19}, numerical \cite{EBW+20}, and experimental evidence \cite{AAB+20_a, AAB+20_b} imply that VQS approaches may be among the best candidates for operational quantum advantages in the near term. In \textbf{Sections \ref{sec:Section-IV-sqp}}, we detail a large number of potential applications for quantum simulation and whether they may yield empirical quantum advantages in the near term.

\subsection{Variational quantum machine learning}
\label{sec:Section-III-vqml}

Prior to the advent of the VQA paradigm, the field of quantum machine learning (QML) grew rapidly following the publication of the quantum linear systems algorithm (QLSA) \cite{HHL08}, which yielded a superpolynomial advantage on a classically easy problem. The QLSA algorithm solves linear systems by matrix inversion, a general technique used by many machine learning algorithms (\textit{e.g.} to invert a feature covariance matrix). Given the generality of the QLSA, early QML research largely sought to improve upon it \cite{Amba12, CJS13, CKS17, WZP18} and leverage it towards the development of other QML algorithms for FTQC hardware. Examples of these include quantum algorithms for support vector machines (SVM) \cite{RML14}, recommendation systems \cite{KP17}, clustering \cite{LMR13a}, principal components analysis \cite{LMR14}, and Gaussian processes \cite{ZFO+18, ZFF19}, among many others.

However, despite these initial successes several challenges (including the input and output constraints discussed in \textbf{Section \ref{sec:Section-II-pcts}}) imply that the QLSA may not be practical in the near term \cite{Aaro15}. Among these is ``dequantization'' \cite{Tang18a}%
\footnote{Dequantization, in general terms, refers to a quantum-inspired development in classical algorithm theory (such as the discovery of a comparably efficient subroutine) that eliminates a previously claimed advantage for a theoretical quantum algorithm.}, the first example of which occurred in 2018 and involved the development of a classical algorithm for efficient $\mathit{l}_2$-norm sampling of approximate matrix products%
\footnote{In particular, assuming the matrix is low-rank \cite{Tang18a, CGL+20}, the classical algorithm could be used to replace a core subroutine of the quantum recommendation system algorithm \cite{KP17}, reducing the advantage from superpolynomial to polynomial.}. Other QML algorithms based on the QLSA were subsequently dequantized, including ones for principal component analysis and supervised clustering \cite{Tang18b}, support vector machines (SVM) \cite{DBH19}, low-rank regression \cite{GLT18, GST20}, semi-definite program solving \cite{CLL+19}, and low-rank Hamiltonian simulation and discriminant analysis \cite{CGL+20}.

Despite the practical challenges for early QML algorithms, the development of variational approaches has broadened QML into a substantially more diverse field with many algorithms that may deliver practical advantages before FTQC is available. In general, QML approaches can be placed into one of four categories on the basis of the type of data and device being used \cite{Woss21}: 

\begin{enumerate}[leftmargin=20pt]

    \item A learning algorithm executed on a \textbf{classical device} with \textbf{classical data}.
    
    \item A learning algorithm executed on a \textbf{classical device} with \textbf{quantum data}.
    
    \item A learning algorithm executed on a \textbf{quantum device} with \textbf{classical data}.
    
    \item A learning algorithm executed on a \textbf{quantum device} with \textbf{quantum data}.
    
\end{enumerate}

Category \textbf{1} represents classical machine learning approaches, including deep learning. In the context of QIS, classical machine learning may find relevance for benchmarking QML algorithms, controlling quantum hardware, and designing quantum experiments \cite{MNK+18_b}. At present, Category \textbf{2} finds primary application in characterizing the state of quantum systems, studying phase transitions, and high-energy physics. Category \textbf{4} may be particularly relevant to domains at the intersection of quantum metrology \cite{TB15} and QML where quantum data is received as input%
\footnote{While the unification of quantum metrology and QML appears unexplored for biological and clinical research, recent work on quantum algorithms for medical imaging represents a movement in this direction \cite{KVL20}.}. Alternatively, a QML algorithm could be used to post-process quantum data generated by a quantum simulation algorithm, such as the VQE.

Variational QML approaches typically belong to categories \textbf{3} and \textbf{4}. These algorithms include variational incarnations of the QLSA \cite{BLC+19, HBR+19, LJL19}, a broad variety of quantum neural networks (QNNs) (\textit{e.g.} variational autoencoders \cite{KVD+18}, generative adversarial networks \cite{DK18, RA19}, continuous variable QNNs \cite{KBA+19}, and reinforcement learning \cite{CYQ+19})%
\footnote{It is interesting that classical neural networks were conceived as inherently non-linear while the linearity of quantum mechanics is well established (indeed, it was shown in 1998 that were quantum mechanics to be non-linear, quantum computers could solve \textrm{NP}-complete problems in general) \cite{AL98}. This apparent incompatibility does not represent a significant challenge, however. Multiple approaches to QNN design allow for non-linearity to be approximated, including continuous-variable QNNs \cite{KBA+19} and ones leveraging ancilla qubits to implement reversible gates that emulate non-linearity \cite{WDK+17, CGA17}.}, and at least one example of a variational approach to quantum unsupervised learning \cite{OMA+17}. For reviews of this area of the QML field, see \cite{DTB+16, DB17}.

Among variational QML approaches, quantum kernel estimation (QKE) \cite{SK19, HCT+19, LAT20} and variational quantum classifiers (VQC) \cite{SK19, LSI+21, HCT+19} for supervised learning are of particular interest. In the case of QKE, the aim is to leverage a quantum circuit that implements a non-linear map between the classical features of each class and corresponding quantum states%
\footnote{It is notable that changing the data encoding strategy (\textit{e.g.} from basis encoding to amplitude encoding) changes the resulting kernel \cite{Schu21}.}. The inner product of the two quantum states, corresponding to a quantum kernel function measuring the distance between the data points, is then sampled from the quantum device for each sample pair. This kernel, which may be intractable to compute classically \cite{HCT+19, PCH+21}, can then be input into a classical machine learning algorithm, such as an SVM, for a potential quantum advantage. In this respect, QKE can be viewed as combining both Categories \textbf{2} and \textbf{3} in a single method. A recent demonstration of this approach on a 67 feature astronomy dataset using 17 qubits provided evidence that QKE can yield comparable performance to a classical SVM (in this instance, with a radial basis kernel) on a practical dataset \cite{PCH+21}. Further, advantages were recently shown to be possible using these methods on engineered data sets \cite{HBM+21}. Whether these advantages can be realized on realistic data sets remains the subject of ongoing research efforts.

Similarly, VQCs aim to learn a non-linear feature map that embeds the classical features into a higher-dimensional quantum feature space (\textit{i.e.} a region of the Hilbert space of the quantum system) and maximizes the class separation therein. Doing so allows for the use of a simple linear classifier and can be viewed as a variational analog to a classical SVM. Crucially, the VQC approach yields a general framework for quantum supervised learning \cite{LSI+21}. In particular, VQCs can be viewed as implementing a two step process of i) embedding classical data into a quantum state and ii) finding the appropriate measurement basis to maximize the class separation. This approach can be further generalized to allow for other fixed circuits, such as amplitude amplification or phase estimation, following the data embedding phase \cite{Schu21}. A key benefit of VQCs is that the quantum circuit directly outputs a classification, which greatly reduces the number of measurements required relative to QKE.

While many variational QML methods lack analytical bounds on their time complexity (unlike many of the earlier QML algorithms), they may nonetheless offer other forms of quantum advantage. In particular, theoretical and empirical evidence exist for improvements in generalization error \cite{ASZ+20, BBF+20, JZL+20, WDL+21}, trainability (\textit{i.e.} with certain constructions, favorable training landscapes with fewer barren plateaus and narrow gorges \cite{ASZ+20, BBF+20, MBS+18, PCW+20, AHC+21, LYD+21}), and sample complexity \cite{BCG+96, SG04, SWL+19}. It is plausible that these types of advantages may lead to novel machine learning applications in biology and medicine. However, recent work has also made clear that the bar for achieving these advantages may be high given the ability of data to empower classical machine learning algorithms \cite{HBM+21}.

Explorations of variational QML approaches for biology and medicine are only now getting under way. They include proof of principle implementations of protein folding with a hybrid deep learning approach leveraging quantum walks on a gate-based superconducting device \cite{CCM21} and the diagnosis of breast cancer from legacy clinical data via QKE \cite{SBS+20}. As larger, more flexible quantum devices are made available, further growth of research into applications of variational QML is expected.

\subsection{Quantum approximate optimization algorithms}
\label{sec:Section-III-qaoa}

The quantum approximate optimization algorithm (QAOA) is a hybrid quantum-classical algorithm for targeting optimization problems, such as MaxCut \cite{FGG14}. Following its original publication, the algorithm was generalized into a broader framework, called the quantum alternating operator ansatz (retaining the original acronym) \cite{HWO+19, HWR+17}. This generalization allows for more expressive constructions capable of addressing a broader range of problems. Below, we briefly review the original QAOA approach; we refer to Hadfield et al. \cite{HWO+19} for more details including extensions to the algorithm.

The original QAOA leverages two Hamiltonians, a phase Hamiltonian, $\mathit{H}_P = f(y) \vert y \rangle$, and a mixing Hamiltonian, $\mathit{H}_M = \sum_{j=1}^n X_j$. $\mathit{H}_P$ encodes the cost function which operates on an $n$ qubit computational basis state and $H_M$ is comprised of a Pauli-X operator for each qubit. Application of the unitary operators generated by $H_P$ and $H_M$ to the initial state is then alternated for $p$ rounds. Here, $p$ represents a crucial parameter of the QAOA algorithm, defining the length of the quantum circuit and the number of its parameters. There are $2p$ parameters of the form $\gamma_1,\beta_1, \gamma_2, \beta_2, \dots \gamma_p,\beta_p$, which control how many iterations of the alternating Hamiltonian operators are applied. The QAOA circuit thus prepares the parameterized quantum state,

$
    \vert \beta, \gamma \rangle = 
    e^{-i \beta _P H_M}e^{-i \gamma_P H_P} \dots
    e^{-i \beta _1 H_M}e^{-i \gamma_1 H_P} \vert s \rangle
$,

which is an unbalanced superposition of computational basis states (typically, $\ket{s}$ is initialized as a balanced superposition). By measuring all qubits in the computational basis, a candidate solution, $y$, is obtained with probability $| \langle y |\beta , \gamma \rangle |^2$. By repeated sampling of the state preparation and measurement steps, the expected value of the cost function over the returned samples can be computed as $\langle f \rangle = \langle \beta,\gamma | H_P| \beta, \gamma \rangle$.

In many respects, the QAOA algorithm and its generalization are similar to the VQA framework described in \textbf{Section \ref{sec:Section-III}}. However, unlike VQAs, QAOA algorithms are less flexible in the problems they can address. This is due to i) limitations on the construction of the problem Hamiltonian and ii) the singular $p$ hyperparameter, which defines how many applications of the problem and mixing Hamiltonians should occur. Together, these characteristics may limit the expressivity of QAOA algorithms and, in turn, their ability to yield a practical quantum advantage. Despite these potential challenges, the QAOA algorithm may have some benefits over other quantum algorithm approaches due to its relatively short circuit depth%
\footnote{This makes the somewhat strong assumption that $p$ grows only logarithmically with respect to the problem instance; in practice, it seems likely that polynomial or even exponential scaling may be required to solve certain problems.}, which aligns with the limited coherence of NISQ devices.

With respect to quantum advantages, recent numerical simulations have estimated that an experimental advantage would require hundreds of noisy qubits \cite{GM19, DHK+20}. Such a NISQ device appears plausible in the near term \cite{ZWC+20} and potential advantages using the QAOA framework are being explored on at least one state-of-the-art superconducting platform \cite{HSN+21}. Further, the QAOA framework may also broaden the scope of possible quantum advantages to improved approximation ratios on hard optimization problems%
\footnote{The QAOA algorithm also motivated the development of a novel classical algorithm \cite{BMO+15}, which improved upon the the quantum lower bound for approximation of constraint satisfaction problems.}. Yet, the exact problems for which practical advantages may exist are at present unclear. Relevant to biology, we know of two recent examples of QAOA-based proof of principles. The first applied QAOA to the task of protein folding in a lattice-based model \cite{FBI18}, while another used QAOA to develop an overlap-layout-consensus approach for \textit{de novo} genomic assembly \cite{SAB20}.

To address the gap around the existence of problems where a quantum advantage may be possible, recent work has involved the development of a framework for searching for QAOA-based advantages \cite{LJJ+20} and a classical machine learning approach for identifying problems that may offer an advantage \cite{MCD20}. These general approaches, designed to aid in the targeting of problems that may yield a quantum advantage, remain relatively unexplored. Other work has focused on the characteristics of quantum information and the challenges they present to optimization in the quantum regime, which has led to insights around the type of structure needed for quantum advantages in low-depth circuits \cite{MHM+20}. As is the case with many quantum algorithms, it is too early to say whether QAOA approaches will provide quantum advantages in practice.

\subsection{Quantum annealing}
\label{sec:Section-III-qa}

Quantum annealing (QA) devices provide an alternative approach to quantum computing with specialized NISQ hardware based on classical simulated annealing that may yield quantum advantages in quantum simulation and optimization. Unlike the many gate-based NISQ devices, quantum annealers provide a form of analog quantum computation based on the adiabatic model \cite{FGG+00, AL18}. Existing commercial quantum annealers are qualitatively different from programmable gate-based devices%
\footnote{Like the programmable devices used in the recent quantum primacy experiments \cite{AAB+19, WBC+21, ZCC+21}, commercial quantum annealers also use a superconducting qubit architecture, with some differences in qubit implementation.}. Most notably, their higher qubit counts (now on the order of $10^4$) and qubit connectivity maps allow for relatively large and complex inputs. However, despite these benefits, a number of drawbacks also exist. These include i) a non-universal model of computation%
\footnote{The computational model implemented by practical quantum annealers is believed to be non-universal, unlike the adiabatic model \cite{AVK+07, JGL10}. This limitation is due to the finite temperature of the hardware during computation; in principle, a quantum annealer operating at zero Kelvin could be universal}, ii) their analog nature, which is expected to limit the ability to perform error correction (although error mitigation appears possible), and iii) the development of a quantum-inspired classical algorithm, simulated quantum annealing (SQA) \cite{CA16}, capable of efficiently simulating quantum tunneling%
\footnote{Quantum tunneling is a uniquely quantum mechanical effect that was thought to be a potential source of advantage for QA devices.}.

Given their early development, large qubit counts, and robust connectivity maps, multiple proof of principle demonstrations targeting bioinformatics and computational biology applications have been developed using quantum annealers. These include ranking and classification of transcription factor binding affinities \cite{LDR+18}, the discovery of biological pathways in cancer from gene-patient mutation data \cite{ADR+19}, cancer subtyping \cite{LGB+19}, the prediction of amino acid side chains and conformations that stabilize a fixed protein backbone (a key procedure in protein design) \cite{MMM+19}, various approaches to protein folding \cite{PTT+08, PDD+12, BIF18, OMS+20, HMF21}, and two recent approaches for \textit{de novo} assembly of genomes \cite{SAB20, BRU+20}.

However, despite many promising proof of principle demonstrations (for example, see \cite{KCR+18}), no clear practical quantum advantage has been shown. The degree of advantage possible by QA algorithms also remains unclear and this ambiguity arguably extends over the long term. The primary factors contributing to this uncertainty are the aforementioned non-universal model of computation and the lack of clarity around the possibility for error correction. Thus, while quantum annealing approaches to optimization may yet bear fruit in the NISQ era \cite{CL21}, a number of barriers remain to be addressed \cite{MHM+20}.

\section{Future prospects in biology and medicine}
\label{sec:Section-IV}

In this section we describe a broad variety of potential applications for quantum algorithms. Our aim is to highlight the breadth of both existing quantum algorithms and the types of problems in biology and medicine that they may address. We leverage the quantum advantage framework described in \textbf{Section \ref{sec:Section-II}} and note when a specific application may admit an empirical quantum advantage in the near- or medium-term (summarized in \textbf{Table \ref{table:table4}}). While some of the applications described are not expected to be feasible in the near term – indeed, in some cases, even an FTQC may not be the most appropriate tool for the target problem – it is our intent for the breadth of potential applications and research directions covered to be valuable to an interdisciplinary audience. As such, when possible, we have sought to provide i) quantum scientists with relevant details and references to develop targeted quantum algorithms for applications in biology and medicine and ii) domain computationalists with information on quantum algorithms relevant to applications in the biology and medicine and their prospects for operational quantum advantages in the near- or medium-term.

\begin{table*}[ht!]
    \centering
    \begin{adjustwidth}{-0.5in}{0.5in}
        \begin{tabular}{
            >{\raggedright\arraybackslash}p{.22\textwidth}
            >{\raggedright\arraybackslash}p{.14\textwidth}
            >{\raggedright\arraybackslash}p{.18\textwidth}
            >{\raggedright\arraybackslash}p{.14\textwidth}
            >{\raggedright\arraybackslash}p{.18\textwidth}
            >{\raggedright\arraybackslash}p{.18\textwidth}}
            \toprule \\
            \textbf{Target\newline Application} &
            \textbf{Experimental\newline Demonstration} & \textbf{Hardware\newline Device} & 
            \textbf{Algorithm\newline Type} & 
            \textbf{Classical\newline Complexity} & \textbf{Expected\newline Advantage} \\
            \\ \midrule \\
            
            Protein Folding and Conformation Simulation \cite{PTT+08, PDD+12, BIF18, OMS+20, HMF21} & \center Yes & Quantum Annealer & Quantum Annealing & Polynomial; heuristic approximation & Unknown, up to polynomial \\
            \vspace{0pt} \\
            
            Molecular Docking Simulation \cite{BFB+19} & \center No & Gaussian Boson Sampler & Sampling & Superpolynomial & Unknown; up to superpolynomial \\
            \vspace{0pt} \\
            
            \textit{De Novo} Assembly \cite{BRU+20, SAB20} & \center Yes & Quantum Annealer; Universal Gate-Based Quantum Device & Quantum Annealing, Optimization & Polynomial; heuristic approximation & Unknown, up to polynomial \\
            \vspace{0pt} \\
            
            Sequence Alignment \cite{Holl00, LMB06, PK19, SAA+19, NN21} & \center No & Universal Gate-Based Quantum Device & Optimization & Polynomial; heuristic approximation & Polynomial \\
            \vspace{0pt} \\
            
            Sequence Matching \cite{Mont17, RV03, RBW+17} & \center No & Universal Gate-Based Quantum Device & QML; Search & Polynomial & Up to super-\newline polynomial \\
            \vspace{0pt} \\
            
            Inference of Phylogenetic Trees \cite{EJ11} & \center No & Universal Gate-Based Quantum Device & Optimization & Superpolynomial & Polynomial \\
            \vspace{0pt} \\
            
            Inference of Biological  Networks \cite{ADR+19, BB19, NUM20} & \center Yes & Quantum Annealer & Optimization & Polynomial and superpolynomial & Polynomial \\
            \vspace{0pt} \\
            
            Transcription Factor Binding Analysis \cite{LDR+18} & \center Yes & Quantum Annealer & Optimization & Polynomial; heuristic approximation & Unknown, up to polynomial \\
            \vspace{0pt} \\
            
            Neural Networks \cite{SSP14, WKS14, VBB17, RBW+17, KBA+19, JZL+20, SZY+20, AHK+20, ASZ+20, BBF+20} & \center Yes & Universal Gate-Based Quantum Device & QML & Polynomial and superpolynomial (\textit{e.g.} Boltzmann machine) & Polynomial, problem specific, varies by measure (\textit{e.g.} see \cite{CRR20, ASZ+20}) \\
            \vspace{0pt} \\
            
            \bottomrule
        \end{tabular}
    \end{adjustwidth}
    \vspace{4pt}
    \caption{\textbf{Selected quantum proof of principles relevant to biology and medicine.} A variety of quantum approaches to addressing computational problems in biology and medicine exist, some of which have been experimentally demonstrated. Expected empirical advantages vary greatly. Among the most promising near term applications are ones that leverage quantum simulation and quantum machine learning techniques, such as quantum neural networks.}
    \label{table:table4}
\end{table*}

\subsection{Simulating quantum physics}
\label{sec:Section-IV-sqp}

Simulating microscopic properties and processes at the atomic level \cite{HBD+19} is a key area of computational biology research. These tasks often require quantum mechanical simulations, which are classically intractable for all but the smallest quantum systems. These inherent limitations mean that most classical approaches are approximations and often provide a mostly qualitative understanding. In contrast, many of these same quantum mechanical simulations are a natural task for quantum computers. Beyond the NISQ era, it is expected that simulations of large quantum systems may be used to predict biochemical properties and behaviors that are not efficiently computable with classical devices \cite{EBW+20, RWS+17, CRO+19}. 

In this section we begin with an overview of applications in or relevant to biology and medicine that may benefit from quantum simulation approaches. We then provide a brief summary of three quantum algorithms central to quantum simulation. Finally, we conclude with a discussion of the potential for empirical quantum advantages in the near term.

\textbf{Ground states, binding energies, and reactions.} Calculating the energy of a molecular system is an ubiquitous task in computational chemistry \cite{Jens17}. An important example (which calculates ground states as one of several subroutines) is protein-ligand docking, where the goal is to calculate the binding energy of a small molecule (\textit{e.g.} a drug) to a target site on a protein or other macro-molecular structure (\textit{e.g.} a receptor domain) \cite{PST17}. Though this problem type is often approximated with classical mechanics, future quantum computers may provide highly accurate predictions for docking.
Similar types of simulations may find use as subroutines for calculating protein-protein interactions and small-molecule properties, like the solubility of drug molecules \cite{LCJ+19, HKP+19}.
In addition, calculating the ground state along different nuclear positions yields reaction coordinates, which are essential for understanding both reactivity \textit{in vivo} and drug synthesis mechanisms \cite{DAZ17}.

\textbf{Molecular dynamics} (MD) simulations involve propagating the equations of motion of a microscopic system, such as a complex containing proteins or DNA \cite{Cram04}. 
In addition to understanding qualitative mechanisms, a core goal is often to calculate quantities, such as diffusion rates \cite{MBW20} and Gibbs free energies \cite{GYA+21}.
To do this, MD often uses parameterized force fields and Newtonian dynamics, although one can accurately treat nuclear quantum effects (such as tunneling and zero-point energy) via path-integral MD methods \cite{Tuck10}, and electronic quantum effects using, for example, Car–Parrinello molecular dynamics \cite{Hutt11}.
In principle, quantum simulation approaches may be used both for time propagation (using quantum algorithms that speed up classical ordinary differential equations) \cite{Arrazola2019_pde_nonhom,
Childs2020_pde} and electronic structure calculations performed at each time step \cite{FOG+21}.



\textbf{Excited states.} Though excited electronic and vibrational \cite{MMS+19, SPT21} 
states are not usually a primary focus in biological processes, they are important for probing microscopic states using spectroscopy \cite{Muka99, Bern20}.
Green fluorescent protein (GFP), for example, is a commonly used marker that allows one to study the expression, localization, and activity of proteins in cells via microscopy \cite{Remi11, LC11}. 
Other artificial dyes and markers have also been used to study dynamic processes, such as diffusion \cite{SBH15} and DNA unraveling \cite{KSW+18}.
In principle, the ability to accurately compute excited states could lead to effective screening methods to aid the development of novel fluorescent proteins and dyes that emit or absorb highly specific wavelengths, have narrower emission/absorption bands, or exhibit higher quantum efficiency. These dye markers may also be probed by a variety of spectroscopy methods, such as absorption, emission, and Raman spectroscopy%
\footnote{Each method has different advantages with respect to molecular specificity and spatiotemporal resolution and the appropriate form of spectroscopy varies by application.} \cite{Muka99, Bern20, LC11}.
In addition, time-dependent femtosecond spectroscopy is often necessary when studying certain biomolecular processes and the ability to model femtosecond excited-state behavior could allow for more accurate interpretation of certain experiments \cite{SCG+19}.
Finally, other excited-state processes inherent to biological systems exist that may benefit from quantum approaches, such as photosynthesis \cite{SCG+19} and modeling simple tissue degradation via ultraviolet light \cite{LG19}.


\textbf{Electronic dynamics.} Deeper understanding of some biologically relevant processes might be achieved from the simulation of electron dynamics\footnote{Simulation of electron dynamics is distinct from MD, which is concerned primarily with the motion of the atomic coordinates.}. Cases where one may need to directly simulate the dynamics of electrons include enzymatically-driven reactions such as nitrogen fixation \cite{RWS+17},
biomolecular signaling \cite{HL15_a}, 
biological processes involving radical reactions \cite{Whit05, SSN+14}, 
component processes of neurons and synapses \cite{RCV+13}, 
photosynthetic processes \cite{SCG+19}, 
and the interpretation of electronic behavior in femtosecond spectroscopy experiments, as mentioned above.
Some related fundamental phenomena might also be better understood via direct simulation. One notable example is proton-coupled electron transfer (PCET) \cite{HM07},
which is ubiquitous but only partially understood. The applications for knowledge generated from simulations of electronic dynamics vary widely. One future possibility could be the design of novel enzymes for the development of more sensitive and specific diagnostic assays or novel therapeutics \cite{SKM+16}.



\textbf{Hybrid quantum-classical models.} When modeling a large biomolecular system, researchers sometimes implement a model that uses different approximations for different portions of the system. For example, one might use Newtonian molecular dynamics for the majority of a protein, but perform electronic structure calculations for the protein's reaction site. Another example is to use density functional theory (DFT) \cite{GDL03} as a lower accuracy method and dynamical mean-field theory (DMFT) \cite{KSH+06} as a higher accuracy method for a subsystem of interest \cite{bauer16_hybrid}. Currently, classical examples of such multi-layered approaches exist, such as the ONIOM method (Own N-layer Integrated molecular Orbital molecular Mechanics) \cite{CSR+15}. In principle these existing classical approaches could be modified to run a classically intractable portion on a quantum computer, leaving the rest to run on a classical computer. Already, work on quantum algorithms in this direction is being pursued \cite{bauer16_hybrid,ma20_galli}.


\textbf{Quantum algorithms for quantum simulation.} Arguably, there are three broad quantum computational methods used when studying quantum physical systems: time propagation \cite{NC00, Lloy96}, quantum phase estimation (QPE) \cite{Kita95, KSV02}, and the variational quantum eigensolver (VQE) \cite{MRB+16, PMS+14, FPG+21}. We describe the basic versions of these algorithms below.
%

\textit{Time propagation.} When one is interested in propagating the dynamics of a system, the goal is to approximate the time-propagation operator

\begin{equation}
    |\psi_0(t)\rangle = U(t)|\psi_0\rangle = \exp(-itH)|\psi_0\rangle ,
\end{equation}

where $H$ is the Hamiltonian describing the system of interest. For near term hardware, this is most easily performed using low-order Suzuki-Trotter decompositions \cite{Suzu76, Suzu90}, 
though asymptotically more efficient algorithms also exist for fault tolerant devices \cite{Berry2015, qsp2017, low2019_qubitization, berry15_blackbox, berry15_hamsim_qwalk, childs19_theoryoftrotter,SSJM21}. 
Note also that QPE, discussed next, uses $U(t)$ as a subroutine.

\textit{Phase estimation.} For an arbitrary Hamiltonian, the quantum phase estimation (QPE) algorithm outputs the phase ($e^{-iE_i\tau}$) of the eigenenergy $E_i$ for arbitrary $\tau$, given the input of an eigenvector $|\psi_i\rangle$,

\begin{equation}
    |\psi_i\rangle|0\rangle \xrightarrow{\rm{QPE}} |\psi_i\rangle|e^{-i E_i \tau}\rangle.
\end{equation}

When the input is a mix of eigenvalues, the probability of measuring a particular eigenvalue (eigenphase) is proportional to its overlap-squared. Assuming one has a FTQC and a method for preparing an eigenstate of interest (for example a molecular ground state), the quantum phase estimation algorithm can be used to output the eigenenergy. One can readily determine the eigenvalue $E_i$ from $e^{-i E_i \tau}$, whose precision depends on the number of additional qubits in the second quantum register. 

\textit{Variational quantum eigensolver.} For early generations of quantum hardware, it is likely that approaches based on the VQE will be the only viable option \footnote{We also note that a distinct algorithm called quantum imaginary time evolution (QITE) was recently proposed, which may also be promising for near-term hardware \cite{MST+19}.}. In this method, the goal is to minimize the function
\begin{equation}
    \min_{\vec \theta} \langle \psi(\vec \theta) | H_{\rm{sim}} | \psi(\vec \theta) \rangle
\end{equation}
by varying the parameters, $\vec \theta$. These parameters determine the behavior of the quantum circuit, which prepares the quantum state $|\psi(\vec \theta)\rangle$. Usually these parameters simply control rotation angles for one- and two-qubit gates. A recent review of the VQE \cite{FPG+21} discusses many of the extensions to the algorithm that have been proposed.

In many cases, these algorithms – time propagation, QPE, and VQE – are extensively modified. For example, a variety of strategies to enhance their capabilities have been leveraged in experiments. Among these strategies are 
error mitigation \cite{WCA+21, TBG17, LB17, MYB19, SBS+19, BW20, AAB+20_a, BGC+21, BGK+21, TEM+21, NKS+21},
post-processing to improve accuracy \cite{RGI18, ECB+21}, 
approaches to reducing the number of quantum circuit evaluations \cite{IYL+19, ZTK+20, HKP20},
and approaches to dynamically modify the quantum circuits \cite{GEB+19}.
It is anticipated that these types of practical enhancements, among others, will be crucial to realizing many empirical quantum advantages in the near term both within and outside the space of quantum simulation problems.

\textbf{Prospects for quantum simulation.} Quantum simulation offers some of the strongest prospects for practical quantum advantages. Example applications include finding ground \cite{KMT+17, MPJ+19} and excited states \cite{MRB+16, OKC+19} of electronic degrees of freedom, vibrational degrees of freedom \cite{MMS+19, SPT21}, and more complex degrees of freedom \cite{SMK+20, HHM+21, HGP+15, SH19}, or dispersion interaction between drug molecules and proteins  \cite{CFF+20}.
VQEs in particular have the strong possibility of near term advantages as they scale. In this respect, one promising direction is divide-and-conquer approaches \cite{FMM+20, MFF+21}, which combine multiple VQEs by hierarchical methods to simulate molecules that would otherwise be too large to input into current NISQ hardware. Provided these approaches prove practical, it is possible that simulating larger, biologically relevant molecules – proteins, nucleic acids, drugs, and metabolites – will be feasible in the near term.

Further, potential also exists for other hybrid quantum-classical models. Already, quantum algorithms for embedding models have been developed \cite{bauer16_hybrid, ma20_galli}. Similar algorithms may allow for the treatment of a subsystem (such as a protein active site) with the quantum computer while the rest of the system (such as the solvent and protein) is simulated with a classical computer. Like the VQE and short-duration quantum dynamics, these hybrid approaches may yield empirical quantum advantages in the near to medium term.

In principle, quantum simulation approaches may also be used both for time propagation and electronic structure calculations, which could be performed at each time step. It is possible that short-time quantum dynamics simulations (using quantum algorithms that speed up classical ordinary differential equations, discussed in the next section) \cite{Arrazola2019_pde_nonhom, Childs2020_pde} on near term devices may find limited use in elucidating reaction mechanisms and approximating free energies that remain classically intractable. 

Over the long term, it is possible that quantum simulations may expand to include QPE on FTQC devices. These methods would allow for the precise quantification of biochemical properties and behaviors far beyond the capabilities of classical HPC systems \cite{RWS+17, CRO+19, EBW+20}.
Further, it is also possible that accurate, long duration \textit{ab initio} MD simulations may also be attainable with these methods over the long term.

\subsection{Simulating classical physics}
\label{sec:Section-IV-scp}

Simulations of biological processes governed by classical physics often require searching over a parameter space to optimize a set of classical variables, for which search, optimization, and machine learning algorithms may be used. In addition, the simulation of Newtonian physics and other non-quantum processes is widely used in biological research. For these simulations, ordinary (ODE) and partial differential equations (PDE) are particularly important and have broad applications – from simulating fluid and tissue mechanics to molecular dynamics. Below, we first summarize this application space, briefly review relevant quantum algorithms, and then consider the prospects for empirical quantum advantages.


\textbf{Conformation search.}
It is often necessary to search a large conformational space in order to find a global or near-global optimum \cite{LFH+17, Hawk17, WCR+19}.
Such a search is usually performed over a domain of classical variables (\textit{e.g.} Cartesian atomic coordinates), and may be done in concert with a quantum mechanical method for calculating the energy at each given conformation (as noted above). 
An important example of conformation search in biology is protein folding \cite{LFH+17}, 
but large conformation spaces are also encountered when studying other biomolecules, such as RNA \cite{PSB+10}, 
determining a drug's molecular crystal structure \cite{Bera15},
or identifying pathways in complex reaction mechanisms \cite{CZC+15}.

\textbf{Fluid mechanics.} Many biological processes are governed by fluid mechanics, requiring the simulation of Navier-Stokes equations \cite{RRS+16}. 
Relevant macroscopic processes in this area include the simulation of cardiovascular blood flow \cite{MBS+10} and air flow in lungs \cite{WL10}, as well as some aspects of gastroenterology \cite{MOM+07}. 
On a smaller scale, one may want to simulate highly viscous flow inside or around microbes \cite{SA09}, or capillary flow \cite{JZM+09}. The latter is especially relevant to angiogenesis \cite{SS05} (a hallmark of cancer \cite{HW00, HW11}) and understanding tumor formation and drug permeability \cite{SC13}.
Further, fluid simulations may be use to model designs for chemical reactors and bioreactors,
which are often critical components in drug \cite{HR14} or complex tissue \cite{HS08} manufacture.


\textbf{Non-fluidic continuum mechanics.} Macroscopic modeling is important not just for fluids but also for solid or semi-solid continuum materials. In this context, the finite element method and related approaches are often used \cite{Hugh12, LLF+06}. 
While applications of these methods include the modeling of macroscopic tissues, such as muscle \cite{SPB+19} and bone \cite{PMC+17}, they have also been applied to the nanoscale, including in simulations of the cytoskeleton \cite{PSJ+14}.

\textbf{Classical electrodynamics.} Classical electrodynamics can ultimately be described by differential equations, too. The design of medical devices is one area where this type of simulation may be useful. Examples may include the modeling of MRI designs \cite{ZG11} or, perhaps, medical devices that interact with lasers.
Another area of application is the design of classical optical devices \cite{GCG20}, which may be used in biological research \cite{WVH+20}.

\textbf{Systems modeling and dynamics.} There are many ubiquitous classical modeling approaches apart from those that directly use Newtonian physics. A particularly relevant example exists in complex population models, which are essential in fields such as epidemiology \cite{BCF19, SK18}. 
Others include detailed simulations of entire cells \cite{BB20}, 
organs \cite{CWZ+21}, 
or groups of organisms \cite{AA20, HRO+15}.

\textbf{Quantum algorithms for classical simulation.}
A number of targeted quantum algorithms have been proposed for finding low-energy conformations and searching through candidate molecules, many of which have been specifically developed for protein folding \cite{LFH+17, WCR+19, BGO+20}. 
More generally, amplitude amplification \cite{BHM+02} may be used to explore conformation spaces over classical variables with a quadratic advantage. Additionally, theoretical quantum advantages have also been shown for other optimization-related subroutines, such as escaping saddle points in optimization landscapes \cite{ZLL20}.
Adjacent to optimization and search, QML algorithms may also be applied. Already, examples exist, such as one for leveraging quantum deep learning to search the chemical space \cite{LTG21, ROA17}. 
It is plausible that empirical advantages with these methods may be achievable in the near term given their hybrid quantum-classical structure. Finally, the past few years have also seen progress in quantum algorithms for solving classical differential equations, either for general cases \cite{Arrazola2019_pde_nonhom, Childs2020_pde, Montanaro2015_mc, Childs2019_spectral_diffeq, Lloyd2020_nonlin} or specific applications, like the finite element method (FEM) \cite{montanaro16_fem, Budinski2021_advec} or Navier-Stokes \cite{Gaitan2020_navier, Budinski2021_navier}. Importantly, among these quantum algorithms are ones for solving the more difficult cases of non-homogeneous and nonlinear PDEs \cite{Childs2020_pde, Lloyd2020_nonlin}. 


\textbf{Prospects for classical simulation.} With respect to search over conformation spaces, it is possible that empirical quantum advantages may be achievable. However, recent work has indicated that general approaches offering quadratic advantages, like amplitude amplification, may be of limited value on their own, even in the FTQC regime \cite{BMG+20}. For this reason, it is expected that the integration of domain knowledge and additional quantum subroutines will be key to achieving any future advantages. For QML, and hybrid quantum-classical algorithms in particular, further exploration of near term compatible methods to improve simulations of classical physics is merited. Finally, the two primary aspects of fluid simulations that lead to simulation difficulty are, arguably, system size and turbulent flow \cite{Sree99}. While it is unclear whether turbulence may be addressed efficiently with quantum approaches, quantum algorithms for differential equations may allow for reductions in complexity with respect to system size \cite{Childs2020_pde, Lloyd2020_nonlin}, which could lead to superpolynomial advantages in some cases. However, a caveat also exists with known quantum algorithms for differential equations given that they are affected by the input and output problems described in \textbf{Section \ref{sec:Section-II-pcts}}. For this reason, further research is required to understand when empirical quantum advantages for these applications may become feasible.



\subsection{Bioinformatics}
\label{sec:Section-IV-opt}

Optimization is central to many bioinformatics tasks, such as sequence alignment, \textit{de novo} assembly, and phylogenetic tree inference. At their core, classical algorithms for these problems often use subroutines for matching substrings, constructing and traversing string graphs, and sampling and counting $\mathit{k}$-mers (\textit{i.e.} substrings of biological sequences). Here, we describe the basic constructions of these problems and summarize relevant quantum algorithms. 

\textbf{Sequence alignment} is a computational primitive of bioinformatics tasks. The heavy integration of sequence alignment algorithms into bioinformatics software has led to diverse applications – from the \textit{de novo} assembly of whole genomes \cite{SN16}, to the discovery of quantitative trait loci linked to disease phenotypes \cite{Doer02} and identification and analysis of driver mutations in cancer \cite{CGK+20}. A classic formulation for identifying the global optimum alignment of two sequences involves finding the lowest weight path through an $n \times m$ dynamic programming matrix, where $n$ and $m$ are the lengths of the sequences being compared (it is often the case that $n = m$) \cite{NW70}. In practice, an approximate solution is typically constructed by a greedy heuristic using a biologically-informed scoring function. Examples of scoring functions include sum-of-pairs, weighted sum-of-pairs, and minimum entropy (each of which imply certain biological assumptions) \cite{Goto99}. For example, with a weighted sum-of-pairs, one may assign different scores to DNA base matches, mismatches, substitutions, insertions, and deletions (the scoring system may also be used to control whether the output alignment is global \cite{NW70} or local \cite{SW81}). Alternatively, for proteins, a scoring matrix may be used where each cell represents the likelihood that the amino acid in the row will be replaced by the amino acid in the column  \cite{Moun08}. This likelihood may be determined empirically by a statistical analysis of a large protein sequence database or on the basis of chemical properties of the amino acids (\textit{e.g.} polar or non-polar, hydrophilic or hydrophobic). While pairwise alignment has polynomial complexity, the generalization to multiple sequence alignment (MSA) with sum-of-pairs scoring is known to be \textrm{NP}-hard \cite{WJ94, BV01, Just01}. Other heuristic approaches to sequence alignment also exist, such as progressive alignment \cite{FD96}. For more on MSA algorithms and their broad applications, see this recent review \cite{CMC+16}.

\textbf{\textit{De novo} assembly} refers to the process of assembling a reference genome – a foundational resource for many bioinformatics analyses – from a large set of overlapping reads of the genome. Often these are short reads (on the order of $10^3$ base pairs long with error rates on the order of $10^{-5}$ \cite{PGB+18}). However, more recent long read sequencing technologies (typically on the order of $\ge 10^5$ base pairs with error rates on the order of $10^{-3}$ \cite{LVE20}) can also be used to aid in the scaffolding of the genome using a hybrid approach \cite{SN16}%
\footnote{A trade-off between read length and base error rates with short and long read sequencing technologies yields complementary characteristics that makes hybrid approaches to \textit{de novo} assembly advantageous (this can be viewed as a close sibling of the well-documented trade-off between depth and coverage in short read sequencing data \cite{SSI+14}). Long read technologies allow for the scaffolding of homopolymeric and repetitive regions that are beyond the length of short read sequences; short read technologies generate many overlapping reads of the sequenced regions providing a large sample size (the depth of coverage for modern short read sequencers ranges from $3\times10^1$ to on the order of $10^4$) and greater statistical confidence for each base call (which can each be treated as an individual hypothesis test).}. Modern software packages typically leverage one of two approaches: i) Overlap-layout-consensus (OLC), which involves the construction of a string overlap graph and is reducible to the \textrm{NP}-complete Hamiltonian path problem (a greedy approximation heuristic is used) and ii) a $k$-mer graph approach, which involves the construction of a de Bruijn graph and is reducible to the Eulerian path problem, which admits a polynomial time algorithm. In practice, achieving high-quality, biologically plausible assemblies is non-trivial and subject to many challenges due to both genome structures, such as homopolymeric and repetitive regions, and the introduction of errors into sequencing data from library preparation, systematic platform error, and low coverage regions \cite{LCM+12}.

\textbf{Phylogenetic tree inference} is the process of inferring the evolutionary relationships between multiple genome sequences%
\footnote{It is notable that phylogenetic tree inference has taken center stage in the COVID-19 pandemic with tools such as Nextstrain \cite{HMB+18}, which has been used to monitor the evolution of SARS-nCoV-2.}. Typically, inference of a phylogeny involves combining i) a multiple sequence alignment of the genomes in question, ii) an evolutionary dynamics model accounting for the types of evolutionary processes occurring between sequences, and iii) a tree's topology, where branch lengths represent the distance (\textit{e.g.} Hamming or Levenshtein distance) between two sequences. The evolutionary dynamics model may be as simple as a continuous-time Markov model sampling from a table of specific events with empirically estimated transition probabilities (\textit{e.g.} a base substitution $\mathrm{T}>\mathrm{A}$; in evolutionary biology, the events and their probabilities may be highly specific to a species or genus). Alternatively, more complex Bayesian methods and mixture models may be used where the associated probabilities for events may vary for different sequence regions. The topology of a phylogenetic tree may be initialized randomly from the MSA and is often inferred by hierarchical clustering with a maximum likelihood estimator (MLE) \cite{AAM+21}. In practice, phylogenetic tree methods are central to evolutionary biology \cite{KYT20} and understanding the evolutionary dynamics of clonal populations in cancer \cite{SWK+17} (among a multitude of other applications), the latter of which has significant clinical relevance to the targeting of precision therapeutics and characterization of treatment resistance.

\textbf{Emerging application areas.} Many emerging application areas in bioinformatics exist that may represent interesting targets for quantum algorithm development. Examples of application areas include i) the inference of topologically associating domains (TADs; interacting regions of chromosomes governed by the 3-dimensional bundling structure of chromatin in cell nuclei), which are crucial to our understanding of epigenetic mechanisms \cite{PRD+14}, ii) single-cell multiomics, a set of novel methods for generating multi-modal data from single-cell sequencing assays, which allow for simultaneous measurement of a cell's state across the biological layers (\textit{e.g.} genomic, transcriptomic, and proteomic) \cite{SS19}, and iii) improving the modeling and inference of biological networks (\textit{e.g.} interaction networks for genes, transcripts, or proteins). With respect to the latter, this may include the alignment of multimodal networks \cite{GJF+18, XGR+19, VTN+19}, which can be used for predicting the associations between biological and disease processes \cite{XGR+19, VTN+19}, and the modeling of gene regulator networks using complex-valued ODEs \cite{YBZ+21}, which may be especially well suited to quantum information. Given the breadth of the bioinformatics space, these applications represent a very small subset of the potential emerging application space for quantum algorithm development.

\textbf{Prospects for bioinformatics.} A small number of quantum algorithms for problems in bioinformatics have been proposed (\textbf{Table \ref{table:table4}}). These include theoretical algorithms developed for FTQC devices that target \textrm{NP}-hard problems, such as sequence alignment \cite{Holl00, LMB06, PK19} and the inference of phylogenetic trees \cite{EJ11}, which leverage amplitude amplification and quantum walks \cite{Vene12}. To be made practical, these theoretical quantum algorithms are expected to require both significant refinement and effort in translation. In the near term, these refinements could include i) recasting them for NISQ devices using the VQA, QAOA, or QA frameworks and ii) integrating greater biological context. Already, examples of this type of work exist for \textit{de novo} assembly \cite{SAB20, BRU+20}, sequence alignment \cite{SAA+19}, and the inference of biological networks \cite{BB19, NUM20}. Over the long term, operational advantages may be pursued by optimizing near term approaches and integrating fast quantum algorithm subroutines where possible. Known quantum algorithms that may be relevant to this work include ones for backtracking \cite{Mont18}, dynamic programming \cite{ABI+18, Rona19}, operating on strings \cite{Mont17, RV03, NN21}, and differential equations \cite{Arrazola2019_pde_nonhom, Childs2020_pde, Montanaro2015_mc, Childs2019_spectral_diffeq, Lloyd2020_nonlin}.

Taking these measures into account, operational advantages for these problems may nonetheless remain among the most difficult to achieve. This is partly due to the factors discussed in \textbf{Section \ref{sec:Section-II-pcts}}. Other barriers to quantum advantages include i) the sophistication of existing classical heuristic algorithms and the inherent parallelism of many of the problems they solve, ii) the scale of both existing classical hardware and practical problem instances within the context of contemporary research \cite{KGC+20}, iii) the broad institutional support and incumbent advantage benefiting existing classical approaches (including extensive clinical validation in the medical setting), and iv) the likely precondition of FTQC to realize polynomial advantages based on amplitude amplification in practice \cite{BMG+20}. Thus, while current research in this direction shows long term promise and should be explored further, many of these quantum advantages appear unlikely to be practical the near term.

\subsection{Quantum machine learning}
\label{sec:Section-IV-qml}

Many of the quantum advantages associated with near term variational QML algorithms relate to model capacity, expressivity, and sample efficiency. In particular, variational QML algorithms may yield reductions in the number of required trainable parameters \cite{SBS+20}, generalization error \cite{ASZ+20, BBF+20, JZL+20, WDL+21}, the number of examples required to learn a model \cite{SWL+19, LAT20}, and improvements in training landscapes \cite{WKS14, ASZ+20, BBF+20, PCW+20, AHC+21}. Evidence supporting one or more of these advantages has been found in both theoretical models and proof of principle implementations of quantum neural networks (QNNs) \cite{ASZ+20, BBF+20, JZL+20, PCW+20} and quantum kernel methods (QKM) \cite{LAT20, WDL+21, HBM+21}. It is notable that these methods are both closely related to VQAs leveraging gradient-based classical optimizers (indeed, they often share overlapping definitions in the literature, as briefly noted in 2019 \cite{KBA+19}) \cite{Schu21, LBN+17, JGH18}. Given the breadth of applications for machine learning approaches in biology, we focus our discussion below on these types of advantages and their potential applications in lieu of specific methods.

\textbf{Improvements to training landscapes} refer to the reduction or removal of barren plateaus and narrow gorges in the landscape of the objective function of a gradient-based learning algorithm. These improvements may stem from the unitary property of (many) quantum circuits, which inherently maintains the length of the input feature vector throughout the computation \cite{SBS+20} provided an appropriate input encoding is used. This bears similarity to many classical approaches used to improve and stabilize training landscapes in practice, such as batch normalization \cite{IS15} and self-normalizing neural networks \cite{KUM+17}. While improved training landscapes may result in more rapid convergence, it is unclear whether this type of advantage alone can be made practical (\textit{e.g.} by allowing for a model to be trained that would be ``untrainable'' by classical means). Fortunately, improvements in training landscapes have been seen to co-occur with reductions in generalization error \cite{ASZ+20, SMM+20}.

\textbf{Reductions in generalization error.} Generalization error measures the ability of a machine learning model to maintain similar performance (\textit{i.e.} "generalizability") on unseen data%
\footnote{It is necessary to assume that the unseen dataset is i.i.d. to the data used to train the machine learning model.}. Reductions in generalization error may yield advantages in the accuracy and flexibility of trained machine learning models. Advantages in generalization error are dependent on a variety of factors, including the encoding used (with basis encoding performing particularly poorly \cite{BPP21}) and the availability of data sufficient to train a comparable classical model \cite{HBM+21}. While there is substantial evidence supporting reductions in generalization error \cite{ASZ+20, BBF+20, JZL+20, WDL+21}, evidence of poor generalization performance under certain constructions also exists (\textit{e.g.} see this recent paper \cite{QWD+21}). This may be partly attributable to shallower quantum circuits providing better utility bounds than deeper circuits \cite{DHL+20_b}, which contrasts with classical neural network intuition where increased layer depth is associated with an exponential increase in model expressiveness \cite{RPK+17}. With respect to applications, much like improvements in training landscapes, reductions in generalization error (despite having broad relevance) may alone be insufficient to provide a practical quantum advantage in the near term. 

\textbf{Reductions in sample complexity} may allow for the learning of robust machine learning models from fewer examples. Intuitively, sample complexity (and generalization error) advantages may arise when quantum entanglement enables the modeling of classically intractable correlative structures. If such sample complexity advantages are achievable with classical data, they will likely be problem instance specific \cite{HBM+21}, highly dependent on the distribution of the input data \cite{LAT20, HBM+21}, and are unlikely to be superpolynomial \cite{SG04, AW17}. Nonetheless, polynomial \cite{BCG+96, SG04, SWL+19} or even sub-linear reductions in the number of examples required to build a classifier could provide significant operational advantages%

The source of these operational advantages can be viewed as the result of the typically high cost of sample generation or acquisition in biological research and clinical contexts. This cost can be due to a variety of factors, including wet lab protocol duration, procedural invasiveness, financial cost, or low disease incidence. Some emblematic examples of costly data acquisition in the clinical context include the sampling of bone marrow in leukemias and time-consuming medical imaging of patients with rare neurological diseases.

This significant cost in the acquisition of data contrasts with typical hardware measures of time and space resources, such as clock cycles, gates, (qu)bits, and queries, which are often (or expected to be) very cheap due to their high frequency and scalability%
\footnote{While this may not be true in the NISQ era, it may well be the case for FTQC devices and is certainly the case for classical HPCs.}. Indeed, where a quadratic reduction in the number of function queries (\textit{e.g.} from $n = 10^6$ to $\sqrt{n} = 10^3$) may lead to only millisecond differences in compute time, a similar reduction in the numbers of samples could save months or years in biological sample collection and processing time (to say nothing of the economic considerations). Further, by providing a potentially large improvement in operational outcomes, sample complexity advantages over classical data distributions may also exhibit a resiliency to improvements in classical hardware. For these reasons, the discovery of data distributions for which quantum computers may offer even small reductions in sample complexity could have substantial relevance to the development of QML approaches for many prediction and inference problems in biology and medicine.

\textbf{Privacy advantages.} Since the earliest days of the QIS field, the inherently private nature of quantum information has been a subject of significant interest \cite{BB84}. More recently, theoretical work at the intersection of quantum information and differential privacy has been explored \cite{ZY17, DHL+20_a, SLK21, WCY21}, with potential applications in the healthcare setting \cite{SMT17}. While research in this area is in the very early stages, the potential for differentially private learning on large, open healthcare data resources presents an opportunity for classical machine learning techniques and may be further empowered by quantum advantages in differentially private learning and data sharing over the long term. For an example of one classical approach, see this recent paper \cite{CGS+20}.

\textbf{Prospects for quantum machine learning} Variational quantum machine learning (QML) is expected to provide a methodological toolbox with significant relevance to a wide range of biological research and clinical applications. Substantial numerical and theoretical evidence now points towards a variety of strengths related to variational QML algorithms on NISQ hardware. These include robustness in the presence of device, parameter, feature, and label noise \cite{BBF+20, SBS+20, NBS19, BPR+20}. It is possible that the breadth of applications may be similar to deep learning, a set of highly flexible methodological tools for generative and predictive modeling now widely used in the field \cite{APP+16, OE16, Topo19}.

With respect to quantum advantages, further experimental work is necessary to assess whether the potential advantages in variational QML discussed above can yield operational advantages. Sample complexity advantages in particular could have a great impact\footnote{It is notable that the existence of polynomial advantages in sample complexity is very much an active research area.}. Indeed, if even small polynomial reductions can be demonstrated over data distributions common in biological and clinical research, they may find important applications where examples are rare (\textit{e.g.} due to disease incidence) or sample acquisition is expensive, invasive, or difficult. Examples that fit this criteria include the diagnosis and prognosis of rare phenotypes \cite{FCG+20, SLS+20}, the identification of adverse multi-drug reactions from EHR data \cite{ZPZ+18, CPS+20}, and the diagnosis of cancers \cite{LSG+19} and their subtyping on the basis of clinical outcomes, such as drug sensitivity \cite{BFR21} and disease prognosis \cite{YZB+16}. Altogether, while practical applications of QML advantages remain largely theoretical, their potential to address existing domain constraints provides ample motivation for further research into variational QML approaches.

\subsection{Quantum data structures}
\label{sec:Section-IV-qds}

In bioinformatics and computational biology, non-traditional data structures have long been leveraged by classical algorithms to great effect. For instance, a number of state-of-the-art algorithms for error correcting sequencing data \cite{LBM16} leverage Bloom filters \cite{Bloo70}, a probabilistic data structure related to hash tables. The core benefit of a Bloom filter comes from its ability to trade a low probability of false positive lookups for significant savings in memory – a common constraint in large bioinformatics pipelines. In a similar vein, the full-text minute space (FM)-index data structure \cite{FM00} is leveraged (in conjunction with the Burrows-Wheeler transform) by sequence aligners such as Bowtie \cite{LTP+09}, the BWA family of aligners \cite{LD09, LD10, Li13}, and more recent graph reference genome aligners \cite{PNE+17, RSL+19}. Like Bloom filters, FM-indexes offer rapid querying and significant memory efficiency.

It is conceivable that the inherently probabilistic nature of quantum computers and novel data input modalities offered by quantum information, such as angle and phase encoding, could lead to the development of similarly useful quantum data structures and abstractions in the FTQC regime. QRAM represents one example \cite{GLM08a, GLM08b, AGJ15, HZZ+19, CDE+21}. In the medium term, a concerted effort towards developing an open quantum data structure library may be useful for improving our understanding of the types of quantum approaches that may admit practical advantages over the long term.


\section{Summary}
\label{sec:Section-V}

The landscape of quantum advantages considers the benefits of quantum technologies relative to existing classical alternatives. An advantage is identified by evidence for which the context can be described as theoretical, experimental, or operational. We divide quantum advantages into four classes on the basis of the strength of the quantum advantage (\textit{i.e.} a polynomial or superpolynomial reduction in complexity) and the complexity of the analogous classical algorithm (\textit{i.e.} polynomial or superpolynomial). Each class implies differing prospects for achieving experimental and operational advantages in the near term. 

Among these four classes, superpolynomial advantages on classically hard problems appear to present the most viable path to operational advantages. In biology and medicine, relevant domain problems in this class include ones related to the quantum simulation of biologically relevant molecules – such as small molecules, protein domains, and nucleic acids – and their chemical quantities. This suggests that the fields of drug development, biochemistry, and structural biology may stand to benefit in the near term from targeted proof of principles using hybrid quantum-classical approaches, such as variational quantum simulation.

More speculatively, quantum machine learning algorithms yielding advantages in sample complexity (even small polynomial ones) may yield meaningful advantages in the near to medium term. In particular, this type of advantage may allow for the training of quantum machine learning models that exhibit generalization error rates similar to classical machine learning models but require less input data. Given the pervasive challenges around generating and acquiring biological and clinical samples, reducing sample complexity may provide a basis for significant operational advantages. Further, these advantages may be resilient to improvements in classical hardware capabilities.

Quantum advantages may result from a variety of quantum algorithm paradigms. These include variational quantum simulation, variational quantum machine learning, quantum approximate optimization algorithms, and quantum annealing. Already, variational quantum algorithms have shown particularly promising results. This is partly due to their substantial flexibility, which allows them to tackle a wide variety of problems across quantum simulation, quantum machine learning, and optimization. Notably, while VQAs may yield superpolynomial advantages on classically hard problems, whether these near term algorithms can fully capitalize on the computational power afforded by quantum information remains a matter of theoretical investigation (\textit{e.g.} see \cite{Ansc21}). Nonetheless, quantum primacy experiments \cite{AAB+19, WBC+21} leveraging parameterized quantum circuits, the core quantum component of near term hybrid quantum-classical algorithms, have already demonstrated the viability of quantum advantages on NISQ devices. 

Finally, we note that it remains possible that the greatest fruit of research into quantum approaches will be novel quantum inspired classical algorithms. For example, the previously noted framework \cite{CGL+20} for the ``dequantization'' of QML algorithms \cite{Tang18a, Tang18b} based on the QLSA has led to the development of classical algorithms that may in the future improve upon existing practical implementations. Similarly, the development of one classical optimization algorithm for constraint satisfaction \cite{BMO+15} was inspired by the original QAOA \cite{FGG14} and improved upon its performance. Further, recent work on the value of data in classical computation has extended our understanding around the types of hard problems that may be tractable with high quality data and classical machine learning techniques \cite{HBM+21}. On the hardware side, stiff competition between quantum approaches \cite{AAB+19} and their classical counterparts \cite{PGN+19, VLB+20, CZX+18, HZN+20, LLL+21}, and the potential that decoherence may not be tamed, together leave open the possibility that large quantum advantages or fault tolerant quantum computers may not be possible. However, a diverse and growing body of evidence – from recent work on novel, more efficient error correcting codes \cite{HH21, GNF+21}, to the identification of significant operational quantum advantages in the energy required to perform computations \cite{VLB+20, BDF+21} – gives much reason for optimism.

\section*{Acknowledgements}

We would like to thank Srinivasan Arunachalam, Meysam Asgari, Aaron Cohen, Justin Gottschlich, Jennifer Paykin, Marek Perkowski, Robert Rand, Xubo Song, and Guanming Wu for productive discussions around topics covered in this work. B. Cordier would like to acknowledge funding by National Library of Medicine and National Institute of Environmental Health Sciences (5T15LM007088-27).

\bibliographystyle{unsrt}  
\bibliography{references}
\clearpage


\end{document}